\documentclass[aos]{imsart}

\RequirePackage{amsthm,amsmath,amsfonts,amssymb}
\RequirePackage[numbers]{natbib}

\startlocaldefs
\theoremstyle{plain}

\newtheorem{theorem}{Theorem}[section]
\newtheorem{lemma}[theorem]{Lemma}
\theoremstyle{remark}
\newtheorem{definition}[theorem]{Definition}

\newtheorem*{fact}{Fact}
\usepackage{url}
\usepackage{algorithm}
\usepackage{graphicx}
\usepackage{siunitx}               
\usepackage{algpseudocode}
\usepackage{caption}
\usepackage{subcaption}
\usepackage{amsmath}
\DeclareMathOperator*{\argmax}{\arg\!\max}
\DeclareMathOperator*{\argmin}{\arg\!\min}

\newcommand{\vectordiff}{\vert \setminus}
\newcommand{\vectorconc}{\|}

\newcommand{\set}{vector}
\newcommand{\concatenation}{concatenation}

\newcommand{\deletion}{deletion}
\newcommand{\logtrans}{log-transform}
\newcommand{\ie}{i.e.,~}
\newcommand{\eg}{e.g.,~}
\newcommand{\replacement}{replacement}
\newcommand{\internalreview}[1]{\textcolor{black}{#1}}%
\usepackage{todonotes}

\endlocaldefs

\begin{document}

\begin{frontmatter}
\title{Log-Paradox: Necessary and sufficient conditions for confounding statistically significant pattern reversal under the \logtrans}
\runtitle{Log-Paradox}

\begin{aug}

\author[A]{\fnms{Ben}~\snm{Cardoen}\ead[label=e1]{bcardoen@sfu.ca}\orcid{0000-0001-6871-1165}},
\author[A]{\fnms{Hanene}~\snm{Ben Yedder}\ead[label=e2]{hbenyedd@sfu.ca}\orcid{0000-0001-7930-8507}},
\author[C]{\fnms{Sieun}~\snm{Lee}\ead[label=e3]{sieun.lee@nottingham.ac.uk}\orcid{0000-0002-9959-6005
}},
\author[B]{\fnms{Ivan Robert}~\snm{Nabi}\ead[label=e4]{ivan.robert.nabi@ubc.ca}\orcid{0000-0002-0670-0513}}
\and
\author[A]{\fnms{Ghassan}~\snm{Hamarneh}\ead[label=e5]{hamarneh@sfu.ca}\orcid{0000-0001-5040-7448}}
\address[A]{Department of Computing Science, Simon Fraser University \printead[presep={,\ }]{e1,e2,e5}}

\address[C]{Precision Imaging Beacon, University of Nottingham \printead[presep={,\ }]{e3}}

\address[B]{Life Sciences Institute, University Of British Columbia \printead[presep={,\ }]{e4}}

\end{aug}

\begin{abstract}
The log-transform is a common tool in statistical analysis, reducing the impact of extreme values, compressing the range of reported values for improved visualization, enabling the usage of parametric statistical tests requiring normally distributed data, or enabling linear models on non-linear data.
Practitioners are rarely aware that log-transformed results can reverse findings: a hypothesis test without the transform can show a negative trend, while with the log-transform, it can show a positive trend, both statistically significant.
We derive necessary and sufficient conditions underlying this paradoxical pattern reversal using finite difference notation.
We show that biomedical image quantification is very susceptible to these conditions. 
Using a novel heuristic maximizing the reversal, we show that statistical significance of the paradoxical pattern reversal can be easily induced by changing as little as 5\% of a dataset.
We illustrate how quantifying the sizes of objects in proportional data, especially where object sizes capture underlying creation and destruction dynamics, satisfies the precondition for the paradox.
We discuss recommendations on proper use of the log-transform, discuss methods to explore the underlying patterns robustly, and emphasize that any transformed result should always be accompanied by its non-transformed source equivalent to exclude accidental confounded findings. 
\end{abstract}



\end{frontmatter}


\section{Introduction}
\label{sec1}
\subsection{Motivation}\label{sec:intro}
Applying a log-transform to a single-dimensional positive vector of numbers is a common technique in statistical analysis that is used to reduce the visualized range, minimize the impact of extreme values~\cite{olivier2008logarithmic}, or enable the application of parametric methods that are only defined on normally distributed data~\cite{tukey1977exploratory}.
The log-transform is one of Tukey's `ladder' of transformations, also known as the `bulging rule', referring to the graphical reshaping effect it has on a distribution.
The log and related transforms can be found in introductory statistical courses~\cite{lane2003introduction, peck2015introduction}, with an accessible introduction to the transformations found in~\cite{fink2009faqs}.
The tail of a distribution is the extreme end of its probability density function (pdf). 
One use case where log-transforms are commonly applied are data following a long-tailed distribution~\cite{lane2003introduction, peck2015introduction, fink2009faqs}.
Intuitively, those are distributions wherein elements become rapidly less frequent as they strongly increase in value. 
In other words, as one moves toward the maximum, not just the value increases, but the spacing between the values increases.
Whether empirical data has a long tail is often determined by two factors: outliers and extreme values.
While both are infrequent, have extreme magnitudes, and numerically cannot be distinguished from each other, there is a subtle yet critical difference in their interpretation.

Outliers are, though not all agree on the terminology here, values that either cannot occur or are induced by accident, noise, measurement error, etc.
Their presence is due to corrupted or noisy observation of the `true' data, they cannot appear in the ideal domain, only in the noisy observed domain.
Ideally, the removal of such values does not change the model one wants to fit to the `true' data. 
Their removal is therefore a valid and often necessary pre-processing step.
If the long tail of the observed data is solely due to outliers, removing outliers is preferred to applying a log-transform, given that the `true' data does not have a long tail and would be unduly reshaped.
We refer the interested reader to the survey of Zimek et al.~\cite{zimek2012survey} for a comprehensive overview of outlier detection.

Extreme values, on the other hand, are so important that they have their own mathematical field~\cite{smith1990extreme, de2006extreme} dedicated to the accurate modelling and prediction of extreme value events.
Examples are earthquakes, stock market crashes, epidemics, and so on. 
In contrast to outliers, accurately modelling and predicting extreme values is vital.
\internalreview{In seismographic recordings, for example, an earthquake would be highly infrequent yet have extremely large values.
Treating an earthquake as an outlier instead of an extreme value, and subsequently removing it from the dataset, would severely limit the value of any subsequent analysis on that dataset. 
In this example, instead of removal, a log-transform could be a valid approach.}

Xiao et al.~\cite{xiao2011use} argue that the decision to use the log-transform should depend on a likelihood test: if the data is better explained by additive error and normally distributed sourced data ($X = N(\mu, \sigma) + \epsilon$), then analysis without transformation is preferred.
However, if statistical tests indicate multiplicative error and log-normal data ($X = \exp{(N(\mu, \sigma))} * \epsilon$) is a better fit, then the log-transform and thus geometric mean is preferred.
When models are fit to data, the resulting coefficients are used to draw conclusions on the underlying data generation functions or populations.
Along similar lines, Feng et al.~\cite{feng2013log} report on the potential of misinterpretation of log-transformed results, and argue that the transform should be applied if it is known that the source data is sampled from a log-normal distribution. 
If no such knowledge is available, they argue in favor of using methods such as generalized estimation equations.
However, such methods are not entirely non-parametric, they still rely on estimation of the first two moments of the measured data: the mean and standard deviation.
We will show that reliance on this first moment can induce paradoxical conclusions{\color{black}: the log-paradox. Here, we present conditions that induce this  seemingly confounding pattern reversal introduced by log-transforming long tail data and introduce approaches to both identify, formally describe, and address the log-paradox under several conditions.}
\subsection{Definition}\label{sec:def}
More formally, the log-transform is a special case of the Box-Cox~\cite{box1964analysis} transform of a real scalar number $y$, given a parameter $\lambda$ as:
 \[ f(y) = \begin{cases} 
          \frac{(y^\lambda-1)}{-\lambda} & \lambda\neq 0 \\
          \log(y) & \lambda = 0
       \end{cases},
    \]
where $f(y)$ returns the reciprocal, normal, square root, and log of $y$ for values of, respectively, $\lambda \in $[-1, 1, 1/2, 0].
Note that $\lambda$ is a hyper-parameter that specifies the transformation, not the operand (y). If $y = 0$, a log-transform leads to undefined values. Note that in the remainder of this paper, the expression $Y=\log(X)$ is shorthand for the element wise application of log to the vector $X: Y = \log(X) = [\log(x_i) ~~\forall x_i \in X]$.
In this work, we focus on the log-transform, highlighted by Keene~\cite{keene1995log} as a `special' or favoured transform from the family of commonly used transforms. 
One reason given is that for positive data its range ($-\infty, +\infty$) matches that of its most frequent target distribution, the Normal distribution. 
The transform can be parameterised by a linear offset: 
\begin{equation}
    y(x) = \log(x - c) ~ \forall x \in X, c \in \mathbb{R}^+, c < \min(X),
\label{eqn:minshift}
\end{equation}
where $c$ is a lower bound for $x$ to adjust frequency weighting on small values. 
Alternatively, setting the base $b$ of the logarithm in function of the distribution of $X$ can also be a useful domain-specific choice.
Another variation is clamping: 
\begin{equation}
  y=\log(\max(x, \epsilon)),    ~~~\epsilon > 0,
\label{eqn:maxshift}
\end{equation}
to avoid domain errors in computation without removing zero valued data with $\epsilon$ near machine precision ($\sim 1.0 e-12$).
When $0<x<1$, as is the case for probabilities in Bayesian functions, the variation $y = -\log(x)$ is not uncommon.

Applying the log-transform is not without risk. Feng et al.~\cite{changyong2014log} detail several ways a log-transform can lead to confounding conclusions, reporting that compression of a range of numbers is not always guaranteed and that the log-transform should only be applied if the transformed domain is also interpretable.
An example where the log-transformed domain is interpretable is in measuring sound with scales such as decibel (dB).
In contrast, interpreting log-scaled temperature, velocity, or surface may result in units that are hard to understand, especially given the fact that a log-transform is parameterized by \internalreview{an appropriately chosen base $b$, commonly $e$ or 10. We further discuss the effects of $b$ in Section~\ref{subsec:choiceb}}.

Using in-silico data, Feng et al.~\cite{changyong2014log} illustrate several scenarios under which the transform can lead to hard to interpret findings.
More recently, Gao et al.~\cite{gao2019log} report a corrected way of defining and computing variance and uncertainty under the log-transform, in the context of counting bacteria.
From these works, it should be clear that without a deeper understanding and despite its commonality and utility, the log-transform should be applied \internalreview{with care} and expertise.
\subsection{The link between log-transform and the geometric mean}
Common hypothesis tests compare the expected value of sample or population distributions for (in)equality or ordering.
For discrete data, the expected value is often approximated by the sample arithmetic mean.
If one log-transforms data, however, the arithmetic mean of the transformed data is directly related to the geometric mean~\cite{gao2019log}.
Let $X \in \mathbb{R}^{\vert X \vert}, X = [x_i > 0 \vert \forall x_i \in \mathbb{R}], b > 1, N=\vert X \vert$: 
\begin{equation}
    \mathbb{E}[\log(X)]=\frac{\sum^{N}\log_b(x_i)}{N},
\end{equation}
\begin{equation}
    \mu^*(X)=\sqrt[N]\prod x_i = \exp_b(\frac{\sum\limits_{i=1}^{N}\log_b(x_i)}{N}).
\end{equation}
The notation $\mu^*$ refers to the geometric rather than the arithmetic mean.
The geometric mean is a very powerful metric, commonly used in domains where quantities are expressed in ratios, not absolute scalar values, for example compound interest in finance, though this does not prohibit the use of geometric mean on absolute scalar values.

In their seminal article, Fleming et al.~\cite{fleming1986not} show that the geometric mean can avoid confounded conclusions and pitfalls, especially when comparing benchmark results on computing systems, which are typically standardized data.
Standardization, here, refers to the practice where numerical data is expressed as a ratio, or relative to a common scale or reference distribution.
Examples of `standardized' data are interest rate, or the standard score itself $z=\frac{x-\mu(X)}{\sigma(X)}$.
Fleming et al.~\cite{fleming1986not} argue that when data is expressed in ratios and thus with common scales, the geometric mean is a more appropriate aggregate metric compared to the arithmetic mean.
It is not always obvious that recorded scalar values are implicit ratios. For example, if one measures the sizes (volume) of organelles in a cell, the distribution of those sizes is limited by the size of each cell, as well as other biophysical constraints. In other words, there is an implicit upper limit $M$ to the volume, and so each measured organelle is really a ratio with respect to that limit.

Recently, Vogel et al.~\cite{vogel2022geometric} derived closed-form expressions for the geometric mean for a set of common distributions and discuss the implications of its use, under a number of common long tail distributions.
The decision to use or not use the log-transform is not always clear-cut. 
For example, Proietti et al.~\cite{proietti2013does} showed that in a financial forecast application, the geometric mean was beneficial, but the benefit was conditional on how far in the future the forecast was.
St-Pierre et al.~\cite{st2018count} argue that, while they found no paradoxical results in their comparison on real world datasets, model coefficients can differ dramatically if one applies a log-transform, leading to problems in interpretation.
As a result, they argue for a change in model rather than a change in data for more robust conclusions.
So far, the related work discusses the decision to use the log-transform as an `either-or'. 
However, there are applications where transformed and non-transformed data are used together. 
For example, the `signal flatness' metric~\cite{rao2010adaptive} is one instance where, in the context of noisy image segmentation, the ratio of transformed over non-transformed data is leveraged as an informative metric combining both.
\internalreview{While potentially powerful, the decision to use the geometric mean, and thus the log-transform, is non-trivial for non-expert end users.}
\subsection{Transform in the context of aggregation and resampling}
In reporting statistics on structures captured in biomedical images, aggregation, for example aggregating volume into mean volume, is often necessary to avoid violation of the precondition of independence of sampling (Simpson's paradox~\cite{wagner1982simpson}). 
Consider a case where we have 2D images of cells from two groups that received treatments `C' and `D'. Each image contains $N$ cells, and each cell contains a variable number of objects. 
We want to test the hypothesis that the surface area of objects in 2D images is different conditional on the treatment. 
Suppose the actual biological dynamics support the hypothesis: $\mathbb{E}[\text{surface} \vert \text{treatment=C}] > \mathbb{E}[\text{surface} \vert \text{treatment=D}]$.
Simpson showed that when all observations are reported without first aggregating at the cell level, the resulting distribution can be skewed and the analysis can \textbf{reverse} the hypothesis. 
Furthermore, the inflated count of observations can easily push significance values above thresholds.
Independence of sampling is a frequent precondition of hypothesis tests, but this is not necessarily true when one mixes objects across cells. 
We refer to Wagner et al.~\cite{wagner1982simpson} and Spirtest et al.~\cite{spirtes2000causation} for more in depth coverage and discussion of Simpson's paradox.

One way to correct for this is to aggregate at the cell level, for example by estimating the mean, then the distribution of the aggregates can be reported as an unconfounded approximation. 
Resampling is also often used to compute a stable or more representative aggregate. 
One such approach is `bootstrapping'~\cite{Efron1979}, which, assuming the underlying data has finite variance~\cite{athreya1987bootstrap}, can be shown to converge to the population mean more robustly than the non-resampled statistic.
The necessity of resampling then raises the question of what the exact impact of a log-transform is under resampling.

\subsection{Contribution}

In this work, we introduce and discuss the log-paradox, a paradoxical difference in means, \internalreview{pattern reversal} or pattern inversion of \internalreview{two} vectors conditional on the log-transform: $\mathbb{E}[A] > \mathbb{E}[B]$ \textbf{and} $\mathbb{E}[\log(A)] < \mathbb{E}[\log(B)]$.
\begin{figure}
  \centering
  \includegraphics[width=0.65\linewidth, keepaspectratio]{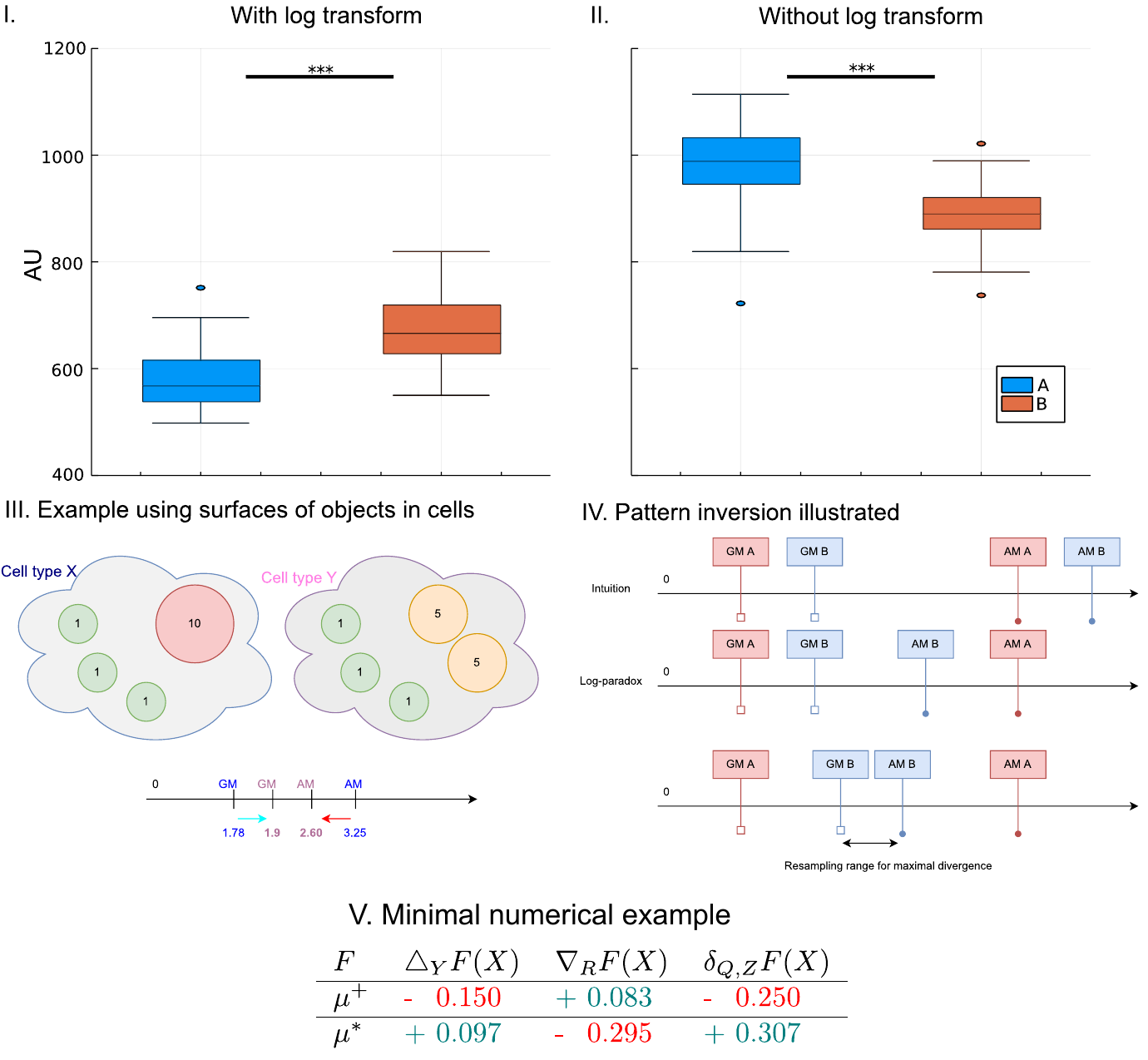}
  \caption{I-II: Log-paradox illustration: \internalreview{A and B are vectors of means sampled from two different in-silico long tail exponential distributions}, seemingly reverse the `is-larger' pattern when log-transform is used before aggregation. \internalreview{A MWU test shows that these sampled means consistently induce a statistically significant reversed pattern.}
  III: Minimal example of paradox in measuring object sizes in biomedical imaging. Two cells, each with a distribution of objects, are quantified by their means. Reporting after a log-transform, would lead to the conclusion that $\mathbb{E}[\log(A)] < \mathbb{E}[\log(B)]$, yet the arithmetic mean shows the inverse trend $\mathbb{E}[A] > \mathbb{E}[B]$. In reality, the underlying pattern is one of aggregation dynamics, where smaller objects either self-organize into larger objects or dis-aggregate into smaller ones. Because of limited space, and 2-3 dimensional reporting of size, such data easily leads to long tail distributions. This underlying characteristic is a sufficient cause for the log paradox to appear. IV: Graphical illustration of the log-paradox on the positive real number line, where for 2 chosen vectors (A, B), the log-transform can lead to a pattern reversal depending on the log-transform.
  \internalreview{V. Table illustrating a minimal numerical example. Let $X = [2, 4, 6, 13], Y=[5.5], R=[6], Q=[3,11], Z=[2,13]$ with $\mu^*(X)=4.99, \mu^+(X)=6.25$. The columns are respectively the change in mean under \concatenation{}, \deletion{} and \replacement{}. The rows denote the arithmetic and geometric means. For each column with very few numbers we can select $Y, Z$ to induce a change that is \textbf{opposite} in sign between geometric ($\mu^*$) and arithmetic ($\mu^+$) means.}
  }
  \label{fig:keyhole}
\end{figure}
\internalreview{To help visualize the paradox and frame the importance of the contribution, we illustrate in Figure~\ref{fig:keyhole} the concept graphically. In panels I-II, we plot boxplots of means of sampled exponential distributions that show that the comparison is \textbf{reversed} conditionally on the log-transform. Panel III illustrates that this can even happen with a very small sample, representative of what can occur in real data due to errors in segmentation or actual long tail effects. Panel IV contrasts the intuition of the expected ordering of the means, with what happens when the paradox is present. Finally, panel V contains a minimal numerical example where, with very small changes to a vector $X$, by \concatenation{} ($\triangle_Y$), \deletion{} ($\nabla_Y$) or \replacement{}($\delta_{Q,Z}$), we can induce a paradoxical change in means.}{\color{black}{The remainder of the manuscript is structured as follows:}}

 \begin{itemize}
     \item We introduce finite difference notation for the arithmetic and geometric mean to rigorously describe this pattern in Section~\ref{sec:analytic}.
     \item Next, we derive necessary and sufficient conditions and a heuristic that maximizes the inversion.
     \item We show in Section~\ref{sec:results} using in silico data when and how the inversion can be induced\internalreview{, with specific attention to the sensitivity to the log-paradox under resampling in Section~\ref{sec:resample}}.
     \item An in silico experiment inspired by real biomedical image analysis highlights how this inversion is very relevant to the field, detailed in Section~\ref{exp:iomed}.
 \end{itemize}
We conclude in Section~\ref{sec:disc} with recommendations on how to approach data analysis when paradoxical inversion is possible.

\section{Method: Analytical expressions for paradoxical mean differences}
\label{sec:analytic}
\subsection{Preliminaries and definitions}
To avoid ambiguity, we first rigorously define the mathematical terms used in this paper.
At the end of this section, we enumerate all terms for ease of reference in Table~\ref{tab:definitions}.
Let $G$ be a sampling function $\mathbb{N}_{>0} \rightarrow \mathbb{R}$. The argument to the function is the index of its sample, e.g., the i-th time the function produces a sample. 
Examples of such functions are random samples from Normal, Poisson, Exponential, or empirical distributions.
The remainder of this paper works on recorded samples of real valued numbers $x$, stored in a vector of random variables $X$:
\begin{equation}
X = [x_1, ... ,x_i, ..., x_N] \in \mathbb{R}^N, x_i = G(i), x_i >= 0, \forall i \in \mathbb{N}_{>0}\label{eqn:vector}.
\end{equation}

\begin{definition}[Vector transform functions]
When a vector of values is transformed, in this paper, an elementwise application of a real valued function $f$ to a vector $X$ is used, denoted by $F(X)$ or $f.(X)$, notation common in vectorized programming languages, let $f: \mathbb{R} \rightarrow \mathbb{R}$:
\begin{equation}
f.(X) = F(X) : \mathbb{R}^{\vert X \vert} \rightarrow \mathbb{R}^{\vert X \vert } =  [f(x_1), ..., f(x_N)] ~\forall x_i \in X,
\label{eqn:vfun}
\end{equation}
where $\vert X \vert = \text{dim}(X)$, the number of elements in the vector $X$.
\end{definition}
For example, $\log_{10}.([10, 100]) = [1, 2]$.
\begin{definition}[Vector concatenation]
Two vectors can be concatenated:
\begin{equation}
    X \vectorconc Y : \mathbb{R}^{\vert X \vert} \times \mathbb{R}^{\vert Y \vert} \rightarrow \mathbb{R}^{\vert X \vert + \vert Y \vert} = [x_1, ..., x_N, y_1, ..., y_M] ~~ \forall n, m >= 1. \label{eqn:vconc}
\end{equation}
\end{definition}
Example: $[1,2,3] \vectorconc [3,4] = [1,2,3,3,4]$.
\begin{definition}[Vector difference]
The difference of two vectors can be defined as the inverse of concatenation:
\begin{equation}
    X \vectordiff Y : \mathbb{R}^{\vert X \vert} \times \mathbb{R}^{\vert Y \vert} \subset X \rightarrow \mathbb{R}^{C=\vert X \vert - \vert Y \vert} = Z = [z_1, ..., z_C] \vert \forall z_i \in X~~ \land Z \vectorconc Y = X. \label{eqn:vdiff}
\end{equation}
\end{definition}
For example:$[1,2,3,3,4] \vectordiff [3,4] = [1,2,3]$.
However, $[1,2,3] \vectordiff [3,4]$ is undefined given that the element 4 does not appear in $X$.
The choice for `vector' here is due to subtleties in interpretation, where `set' or `sequence' or even `series' are sometimes used to denote the same, but they have different semantics. 
A set can be read to imply a cardinality of 1 for a given scalar element, yet here a scalar element can be sampled multiple times. 
A sequence implies ordering, but the operations we apply are invariant to ordering. 
Therefore, in this work, denoting a group of numbers sampled from a generating function by vectors is the most accurate.
\begin{definition}[Aggregating functions]
We define an aggregating function as
\begin{equation}
    f:\mathbb{R}^N \rightarrow \mathbb{R}~~\forall N >= 1.
\end{equation}
\end{definition}
A simple example is the sum of a vector: $f(X) = \sum\limits_{x \in X} x$. 
\begin{definition}[Arithmetic mean]
This neatly brings us to the arithmetic mean:
\begin{equation}
    \mu^+(X):\mathbb{R}^{\vert X \vert} \rightarrow \mathbb{R} : \frac{\sum X}{\vert X \vert }.\label{eqn:am}
\end{equation}
\end{definition}
The $\mu$ symbol denotes `mean', the + superscript is used to reference the arithmetic mean, to differentiate it from the geometric mean.
\begin{definition}[Geometric mean]
\begin{equation}
    \mu^*(A):\mathbb{R}^{\vert X \vert} \rightarrow \mathbb{R} : \sqrt[\vert X \vert]{\prod_{x_i \in X} x_i}  ~~~ \forall x_i > 0.\label{eqn:gm}
\end{equation}
\end{definition}
For reasons of numerical precision, overflow, and performance, the geometric mean is often computed as the arithmetic mean of the log-transformed vector:
\begin{equation}
    \mu^*(A) = \exp_b \log_b \sqrt[\vert A \vert]{\prod_{a_i \in A} A} = \exp_b{\frac{\sum \log_b.(A)}{\vert A \vert}} = \exp_b\mu^+(\log_b.(A)),
\end{equation}.
The impact of the choice of base $b$ is often underestimated, but can be quite dramatic in re-weighting a subset of the distribution, which we discuss in more detail in Section~\ref{subsec:choiceb}.

Note that the geometric mean of a vector collapses to zero if at least one element is zero, and in its most common implementation, using the logarithm, would actually lead to a domain error. 
For the remainder of this paper, we only consider strictly positive vectors $X = [x_1, ..., x_N] ~~ x_i \in \mathbb{R}_{>0} \forall x_i$.
For data with zero entries or negative values, several transformations (Eqn.\ref{eqn:minshift}, \ref{eqn:maxshift}) can be used to avoid domain errors, but this needs careful consideration given that the transformed domain now becomes even harder to interpret.
\begin{fact}[AM-GM inequality]
The geometric and arithmetic mean represent each other's lower and upper bound by the `AM-GM inequality', a specific case of the Cauchy-Schwarz~\cite{steele2004cauchy,sedrakyan2018hm} inequalities. 
That is, for a vector $X \in \mathbb{R}^N_{>0}$:
\begin{equation}
    \mu^*(X) \leq \mu^+(X)\label{eqnamgm}.  
\end{equation}
\end{fact}
\begin{definition}[Inter mean distance (ID)]
We define the inter mean distance as the absolute value of the difference between the means:
\begin{equation}
    ID(X):\mathbb{R}^{\vert X \vert} \rightarrow \mathbb{R}_{>0} : \text{abs}( \mu^+(X) - \mu^*(X) ) = \text{abs}( \mu^*(X) - \mu^+(X) ).\label{eqn:id}
\end{equation}
From the AM-GM inequality (Eq.~\ref{eqnamgm}) we can simplify this to:
\begin{equation}
    ID(X):\mathbb{R}^{\vert X \vert} \rightarrow \mathbb{R}_{>0} : \mu^+(X) - \mu^*(X). 
    \label{eqn:ID}
\end{equation}
\end{definition}
\begin{definition}[Paradoxical comparison]
Finally, we define a paradoxical comparison of 2 \set{}s $A$, $B$ under a transform $f$ as:
\begin{equation}
    \text{Paradoxical comparison}(A, B, f) = \mathbb{E}[A] > \mathbb{E}[B] \land \mathbb{E}[f.(A)] <\mathbb{E}[f.(A)].\label{eqn:paradox}
\end{equation}
\end{definition}
In this paper we focus on the paradoxical comparison of \set{}s where $f=\log$.

\subsection{Modelling changes in the mean using finite difference operators}
In this paper, we study how and when the means of two \set{}s $X$, $Y$ can paradoxically differ.
To describe this analytically, we need to enumerate the operations that transform $X$ into $Y$, then describe how the mean can change over a sequence of such operations.
For scalar valued functions, finite difference operators (\eg forward, backward, and central)~\cite{wilmott1995mathematics} are used to approximate similar differences.
For a scalar real valued function $f:\mathbb{R} \rightarrow \mathbb{R}$ and $h \in \mathbb{R}$ the finite difference operators are defined as follows:

\begin{definition}[Forward difference]
\begin{equation}
    \triangle_h f(x) = f(x + h) - f(x).
\end{equation}
\end{definition}
\begin{definition}[Backward difference]
\begin{equation}
    \nabla_h f(x) = f(x) - f(x - h).
\end{equation}
\end{definition}
\begin{definition}[Replacement difference]
\begin{equation}
    \delta_h f(x) = f(x - h) - f(x + h).
\end{equation}
\end{definition}
For our use case, we are interested in the difference when data is aggregated, that is, represented by a single scalar, such as the mean. 
Let $F$ be an aggregating function such that $F:\mathbb{R^N} \rightarrow \mathbb{R}$ with input vector $X$. 
When $X$ is altered using an operator $G:\mathbb{R}^N \times \mathbb{R}^M : \rightarrow \mathbb{R}^K$ with arguments $X, Y$ to produce a vector $X'$. We want to quantify how $F(X')$ compares to $F(X)$:
\begin{equation}
    F:\mathbb{R^N}\rightarrow\mathbb{R},
    D_{(G, Y)} F(X) = F(G(X,Y)) - F(X).\label{eqn:fd}
\end{equation}
There are three operators we are interested in, adding elements (concatenation), removing elements (deletion), \internalreview{or replacing elements (\replacement{})}.
\subsubsection{Finite difference under concatenation}
\begin{definition}[Finite difference under concatenation]
\begin{equation}
    D_{\vectorconc, Y} F(X) = \triangle_Y F(X) = F(X \vectorconc Y) - F(X).
    \label{eqn:concdif}
\end{equation}
\end{definition}
\begin{definition}[Finite difference under \deletion{}]
\begin{equation}
D_{\vectordiff Y} F(X) = \nabla_Y F(X \vectordiff Y) - F(X)\label{eqn:deldif}.
\end{equation}
\end{definition}
\begin{definition}[Finite difference under \replacement{}]
\begin{equation}
    D_{\vectorconc Y  \vectordiff Z } F(X) = \delta_{Y, Z} F(X) = F((X \vectorconc Y) \vectordiff Z)) - F(X)\label{eqn:repdif}.
\end{equation}
\end{definition}
Before we continue deriving expressions for the finite differences in the arithmetic and geometric means, we summarize the terms defined so far in Table~\ref{tab:definitions}.

\subsubsection{Finite difference of arithmetic mean}
Let $X \in \mathbb{R}^N, Y \in \mathbb{R}^M, Z \in \mathbb{R}^K$, $\vert X \vert = N, \vert Y \vert = M, \vert Z \vert = K$, with $N, M, K \in \mathbb{N}_{>0}$, and $K < N$, $M < N$.
\begin{theorem}[Finite difference of arithmetic mean under \concatenation{}]\label{th:fdamc}
\begin{align}
    \triangle_Y \mu^+(X) = \frac{M}{N+M}(\mu^+(Y) - \mu^+(X)).
    \label{eqn:fwdam1}
\end{align}    
\end{theorem}
\begin{proof}[Proof of theorem~\ref{th:fdamc}]
\begin{align}
    \triangle_Y \mu^+(X) & = \mu^+(X \vectorconc Y) -  \mu^+(X)\nonumber \\
    & = \frac{\sum X + \sum Y}{N+M} - \frac{\sum X}{N}\nonumber \\
    & = \frac{\sum X}{N}\frac{N}{N+M} + \frac{\sum Y}{M}\frac{M}{N+M} - \frac{\sum X}{N} \nonumber\\
    & = \frac{\sum X}{N}\frac{-M}{N+M} + \frac{\sum Y}{M}\frac{M}{N+M} \nonumber\\ 
    & = \mu^+(X)\frac{-M}{N+M} + \mu^+(Y)\frac{M}{N+M}\nonumber \\
    & = \frac{M}{N+M}(\mu^+(Y) - \mu^+(X)).
\end{align}
\end{proof}
As a direct consequence:
\begin{equation}
    \triangle_Y \mu^+(X) < 0 \Leftrightarrow \mu^+(Y) < \mu^+(X)\label{eqn:fwdam}.
\end{equation}
\begin{theorem}[Finite difference under \deletion{}]\label{th:fdamd}
\begin{align}
    \nabla_Y \mu^+(X) & = \frac{M}{N-M}(\mu^+(X)-\mu^+(Y))\label{eqn:backam}.
\end{align}
\end{theorem}
\begin{proof}[Proof of theorem~\ref{th:fdamd}]
    \begin{align}
    \nabla_Y \mu^+(X) & = \mu^+(X \vectordiff Y) -  \mu^+(X)\nonumber \\
    & = \frac{\sum X - \sum Y}{N-M} - \frac{\sum X}{N}\nonumber \\
    & = \frac{\sum X}{N}\frac{N}{N-M} - \frac{\sum Y}{M}\frac{M}{N-M} - \frac{\sum X}{N}\nonumber \\
    & = \frac{\sum X}{N}\frac{M}{N-M} - \frac{\sum Y}{M}\frac{M}{N-M} \nonumber\\ 
    & = \mu^+(X)\frac{M}{N-M} - \mu^+(Y)\frac{M}{N-M} \nonumber \\
    & = \frac{M}{N-M}(\mu^+(X)-\mu^+(Y))
\end{align}
\end{proof}
As a direct consequence:
\begin{equation}
    \nabla_Y \mu^+(X) < 0 \Leftrightarrow \mu^+(X) < \mu^+(Y)\label{eqn:condbackam}.
\end{equation}
These two consequences match the intuition that the mean decreases if we add values of $Y$ to $X$ with a mean ($\mu^+(Y)$) smaller than the mean ($\mu^+(X)$, \concatenation{} difference), or remove values with a mean ($\mu^+(Y)$) greater than the mean ($\mu^+(X)$, \deletion{} difference).
The amount by which the mean is changed is a function of the proportion of data added or removed ($M=\vert Y \vert$) to the original data size ($N=\vert X \vert$).
Another consequence is that to compute the mean when adding or removing elements, one only needs the means of $X$ and $Y$, not the individual elements of $X$ and $Y$.

\begin{theorem}[Finite difference of arithmetic mean under \replacement{}]\label{th:fdamr}
\begin{align}
    \delta_{Y,Z} \mu^+(X) & = \mu^{+}(X)\frac{-(M-K)}{N+M-K} + \mu^{+}(Y)\frac{M}{N+M-K} - \mu^{+}(Z)\frac{K}{N+M-K}.
\end{align}
When we replace elements in $X$, e.g. $M=K$, the expression simplifies to:
\begin{align}
     \delta_{Y,Z} \mu^+(X) & =\frac{M}{N}(\mu^+(Y)-\mu^+(Z))\label{eqn:delam}.
\end{align}
\end{theorem}
\begin{proof}[Proof of theorem~\ref{th:fdamr}]
    \begin{align}
    \delta_{Y,Z} \mu^+(X) & = \mu^+((X \vectorconc Y) \vectordiff Z) -  \mu^+(X)\nonumber \\
    & = \frac{\sum X + \sum Y - \sum Z}{N+M-K} - \frac{\sum X}{N}\nonumber \\
    & = \frac{\sum X}{N}\frac{N}{N+M-K} + \frac{\sum Y}{M}\frac{M}{N+M-K} - \frac{\sum Z}{K}\frac{K}{N+M-K} - \frac{\sum X}{N}\nonumber \\
    & = \frac{\sum X}{N}\frac{N-(N+M-K)}{N+M-K} + \frac{\sum Y}{M}\frac{M}{N+M-K} - \frac{\sum Z}{K}\frac{K}{N+M-K} \nonumber\\
    & = \mu^{+}(X)\frac{-(M-K)}{N+M-K} + \mu^{+}(Y)\frac{M}{N+M-K} - \mu^{+}(Z)\frac{K}{N+M-K}.
\end{align}
\end{proof}
As a direct consequence:
\begin{equation}
    \delta_{Y,Z} \mu^+(X) < 0 \Leftrightarrow \mu^+(Y) < \mu^+(Z). \label{eqn:amrep}
\end{equation}
Note the absence of $\mu^+(X) $ in this expression, the change is only dependent on $Y$ and $Z$, though the magnitude of the change is still proportional ($M/N$).
The \replacement{} difference is related to the `sliding`, `running`, or `windowed` averages, commonly used in estimating trends over variable periods of time, for example, infections during epidemics, arrivals at airports, and so on.  
$\delta_{Y,Z} \mu^+(X)$ then expresses the change in a sliding, running, or windowed average with window size $N$ and step size $K$.
For example, recording a weekly running average, with daily records: $\vert X \vert$=7, $K$=1.
A rapidly spreading epidemic can lead to reporting of running averages on a logarithmic scale, in which case the paradox can be induced.
\subsubsection{Finite difference of geometric mean}
\begin{theorem}[Finite difference of geometric mean under \concatenation{}]\label{th:fdgmc}
\begin{align}
    \triangle_Y \mu^*(X) & = \mu^*(X)(\mu^*(X)^\frac{-M}{N+M} \mu^*(Y)^\frac{M}{N+M} - 1)\label{eqn:fwdgm}. 
\end{align}
\end{theorem}
As a result:
\begin{align}
    \triangle_Y \mu^*(X) & > 0 \Leftrightarrow \nonumber \\
   \mu^*(X)^\frac{-M}{N+M} \mu^*(Y)^\frac{M}{N+M} - 1& > 0
    \Leftrightarrow \nonumber\\
    \mu^*(Y)^\frac{M}{N+M} & > \mu^*(X)^\frac{M}{N+M}   \Leftrightarrow \nonumber\\
    \mu^*(Y) & > \mu^*(X). 
    \label{eqn:condfwdgm}
\end{align}
\begin{proof}[Proof of theorem~\ref{th:fdgmc}]
    \begin{align}
    \triangle_Y \mu^*(X) & = \mu^*(X \vectorconc Y) -  \mu^*(X) \nonumber\\
    & = \sqrt[N+M]{\prod X \prod Y} - \sqrt[N]{\prod X } \nonumber\\
    & = (\sqrt[N]{\prod X })^\frac{N}{N+M} (\sqrt[M]{\prod Y})^\frac{M}{N+M} - \sqrt[N]{\prod X } \nonumber\\
    & = (\mu^*(X))^\frac{N}{N+M} (\mu^*(Y))^\frac{M}{N+M} - \mu^*(X) \nonumber\\
    & = \mu^*(X)(\mu^*(X)^\frac{-M}{N+M} \mu^*(Y)^\frac{M}{N+M} - 1).
\end{align}
\end{proof}
In other words, the geometric mean will increase under concatenation if and only if `the geometric mean` of the added elements is larger.
Note that we specified initially that $\mu^*(X)>0$, so division on both sides of the equation is still valid. 
\begin{theorem}[Finite difference of geometric mean under \deletion{}]\label{th:fdgmd}
The \deletion{} difference is defined as:
\begin{align}
    \nabla_Y \mu^*(X) & = \mu^*(X)^\frac{N}{N-M} \mu^*(Y)^\frac{-M}{N-M} - \mu^*(X)\label{eqn:backgm}.
\end{align}
As a result:
\begin{align}
    \nabla_Y \mu^*(X) & > 0 \Leftrightarrow \nonumber\\
    \mu^*(X)^\frac{N}{N-M} \mu^*(Y)^\frac{-M}{N-M} - \mu^*(X)) & > 0\nonumber\\
    \mu^*(Y)^\frac{-M}{N-M} & > \mu^*(X) \mu^*(X)^\frac{-N}{N-M} \nonumber\\
    \mu^*(Y)^\frac{-M}{N-M} & >  \mu^*(X)^\frac{-M}{N-M} \nonumber\\
    \mu^*(Y) & < \mu^*(X).
    \label{eqn:condbackgm} 
\end{align}
\end{theorem}
In other words, removing values ($Y$) whose mean is smaller than the mean ($\mu^*(X)$), will increase the mean ($ \nabla_Y \mu^*(X) > 0$).
\begin{proof}[Proof of theorem~\ref{th:fdgmd}]
\begin{align}
    \nabla_Y \mu^*(X) & = \mu^*(X \vectordiff Y) -  \mu^*(X) \nonumber\\
    & = \sqrt[N-M]{\frac{\prod X}{\prod Y}} - \sqrt[N]{\prod X } \nonumber\\
    & = \frac{\mu^*(X)^\frac{N}{N-M}}{\mu^*(Y)^\frac{M}{N-M}} - \mu^*(X) \nonumber\\
    & = \mu^*(X)^\frac{N}{N-M} \mu^*(Y)^\frac{-M}{N-M} - \mu^*(X).
    \label{eqn:backgm} 
\end{align}
\end{proof}
\begin{theorem}[Finite difference of geometric mean under \replacement{}]\label{th:fdgmr}
We restrict to the case where the size of $X'$ is kept constant: $\vert Y \vert = \vert Z \vert = M$.
\begin{align}
    \delta_{Y, Z} \mu^*(X) & = \mu^*(X)( \frac{\mu^*(Y)^{M/N}}{\mu^* (Z)^{M/N}} -  1)\label{eqn:delgm}.
\end{align}    
\end{theorem}
Consequently:
\begin{align}
    \delta_{Y,Z} \mu^*(X) & > 0 \Leftrightarrow \nonumber\\
    \frac{\mu^*(Y)}{\mu^*(Z)} & > 1.
    \label{eqn:gmrep}
\end{align}

\begin{proof}[Proof of theorem~\ref{th:fdgmr}]
\begin{align}
    \delta_{Y, Z} \mu^*(X) & = \mu^*((X \vectorconc Y) \vectordiff Z ) -  \mu^*(X)\nonumber \\
    & = \sqrt[N]{\prod X * \prod Y *(\prod Z)^{-1}} -  \sqrt[N]{\prod X}\nonumber \\
    & = \sqrt[N]{\prod X} \sqrt[N]{\frac{\prod Y}{\prod Z}} -  \sqrt[N]{\prod X} \nonumber\\
    & = \sqrt[N]{\prod X}( 
    \sqrt[N]{\frac{\prod Y}{\prod Z}} -  1) \nonumber\\
    & = \mu^*(X)( \frac{\mu^*(Y)^{M/N}}{\mu^* (Z)^{M/N}} -  1).
    \label{eqn:delgm}
\end{align}    
\end{proof}

\subsubsection{Finite differences of intermean distance}
As finite difference operators are applied to $X$, such as concatenation, deletion, replacement, (Eqn. \ref{eqn:concdif},\ref{eqn:deldif},\ref{eqn:repdif}), the inter-mean distance (ID) changes (Eqn.~\ref{eqn:ID}).
These ID differences are easily expressed, by rearranging terms, in terms of the finite differences of the means:
\begin{align}
    \triangle_{Y} ID(X) & = (\mu^+(X \vectorconc Y) - \mu^*(X \vectorconc Y)) - (\mu^+(X) - \mu^*(X)) \nonumber\\
    & = \triangle_Y \mu^+(X) - \triangle_Y \mu^*(X).\label{eqn:fwdid}
\end{align}

\begin{align}
    \nabla_{Y} ID(X) & = (\mu^+(X \vectordiff Y) - \mu^*(X \vectordiff Y)) - (\mu^+(X) - \mu^*(X)) \nonumber\\
    & = \nabla_Y \mu^+(X) - \nabla_Y \mu^*(X).\label{eqn:backid}
\end{align}

\begin{align}
    \delta_{Y, Z} ID(X) & = (\mu^+((X \vectorconc Y) \vectordiff Z) - \mu^*((X \vectorconc Y)\vectordiff Z)) - (\mu^+(X) - \mu^*(X)) \nonumber\\
    & = \delta_{Y,Z} \mu^+(X) - \delta_{Y,Z} \mu^*(X).\label{eqn:delid}
\end{align}

\begin{figure}
  \centering
  \includegraphics[keepaspectratio, width=0.6\linewidth]{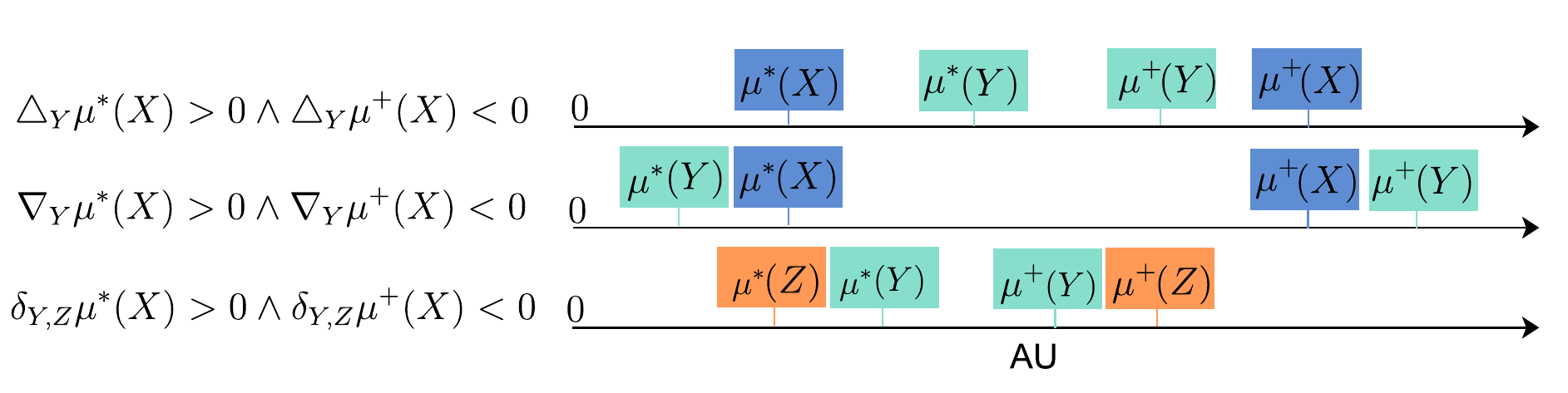}
  \caption{The expressions on the left-hand side list the three conditions for altering $X$ via perturbation arguments $Y$ and $Z$ in order to induce the paradox. The corresponding graphical representations on the right show the relative ordering, along the number line with arbitrary units (AU), of the geometric and arithmetic means of $X$, $Y$, and $Z$ needed to satisfy the conditions.
Note that for the \replacement{} difference $\delta$ the expression is only a function of $Y, Z$, and not of the original vector $X$. Recall that for any vector $A: \mu^*(A)\leq\mu^+(A)$.
These results follow from combining Tables~\ref{tab:sumdiff},~\ref{tab:opposite} and the AM-GM inequality (Eqn.~\ref{eqnamgm}). 
See Table~\ref{tab:definitions} for a listing of all symbols.
}
\label{fig:oppositec}
\end{figure}

Table~\ref{tab:opposite} and Fig.~\ref{fig:oppositec} summarize the necessary conditions for paradoxical change in the means.
These can also be expressed in terms of the inter-mean distance for example, it is true that the condition for the \replacement{} difference then leads to $ID[Y] < ID[Z]$.
However, note that the inter-mean distance is necessary, \internalreview{although} not sufficient; one can trivially find $Y, Z$ such that $ID[Y] < ID[Z]$ and yet the paradoxical difference is not induced.
The intermean distance does not enforce the relative ordering of the means of Y and Z. 
For example, a situation where $\mu^*(Y)<\mu^+(Y)<\mu^*(Z)<\mu^+(Z)$ with $ID[Y] < ID[Z]$ breaks the condition. 

\paragraph{An illustrative example}
We will illustrate how the arithmetic and geometric means change, using the finite difference operators, using a very simple example, \internalreview{quantified in Table~\ref{fig:keyhole}-V.}
Let $X = [2, 4, 6, 13], Y=[5.5], R=[6], Q=[3,11], Z=[2,13]$ with $\mu^*(X)=4.99, \mu^+(X)=6.25$.
This example follows the steps outlined in Algorithms~\ref{alg:insert},~\ref{alg:replace}, selecting elements either in the inter-mean interval or the pair of minimum/maximum, and follow directly from the observations in Table~\ref{tab:opposite}, Fig~\ref{fig:oppositec}.
\internalreview{The minimal example shows easily induced inversely signed differences under the 3 operations: concatenation, difference, and replacement}.
\subsection{Maximizing paradoxical difference in means}
So far, we have derived when a paradoxical comparison, as defined in Eqn.\ref{eqn:paradox}, between two 
\set{}s $X$ and $X'$ can arise (Table~\ref{tab:opposite}).
Here we investigate the magnitude of the paradoxical difference. 
The magnitude directly determines how strongly the paradox can confound common statistical tests.
Given two vectors of real values $X$ and $X'$, two frequent measures of comparing them are hypothesis testing and effect size.
Frequently used hypothesis tests involve a significance threshold that is directly influenced by the difference of $\mathbb{E}[X]-\mathbb{E}[X']$.
Similarly, effect sizes~\cite{ellis2010essential} are often computed based on this difference, scaled by standard deviation. 
Recall that a paradoxical difference between two vectors $X$ and $X'$, given a transformation function $f$, is:
\begin{equation}
\mathbb{E}[X] > \mathbb{E}[X'] ~~ \land ~~ \mathbb{E}[f(X)] < \mathbb{E}[f(X')]. \label{eqnmeandiff}     
\end{equation}
or equivalently:
\begin{equation}
\mathbb{E}['X] - \mathbb{E}[X]  < 0  ~~ \land ~~ \mathbb{E}[f.(X')] - \mathbb{E}[f.(X)] > 0.     
\end{equation}
We study the case where $f$ is the log-transform, so the condition becomes:
\begin{equation}
D_A(X, X') = \mu^+(X') - \mu^+(X)  < 0  ~~ \land ~~  
D_G(X, X') = \mu^*(X') - \mu^*(X)  > 0. \label{eqndadg}    
\end{equation}
In other words, $D_G$ (the increase in the geometric mean as $X$ changes to $X'$) and $D_A$ (the increase in the arithmetic mean as $X$ changes to $X'$) carry opposite signs. 
A simple criterion that maximizes the oppositely signed difference $D_A$ and $D_G$ is: 
\begin{equation}
     \argmax_{X'}~ -(D_A(X, X') \times D_G(X, X')) = -(\mu^+(X')-\mu^+(X)) \times (\mu^*(X')-\mu^*(X)) \label{eqnmax}.
\end{equation}
Note that formulating the criterion using addition ($-D_A + D_G$) allows multiple suboptimal solutions.
A trivial suboptimal solution is given by $D_G=\mu^+(X)-\mu^*(X)$ and $-D_A=0$, here only the geometric mean changes, so no pattern inversion is present for the arithmetic mean and no paradoxical comparison results (Eqn.\ref{eqn:paradox}). 
The multiplicative formulation ensures both terms are jointly maximized.
\begin{theorem}\label{th:max}
The expression that maximizes Eqn.~\ref{eqnmax}, assuming we have total freedom in the selection of $X'$, for any $X$, is: 
\begin{equation}
\mu^*(X') = \mu^+(X') = \frac{\mu^*(X) + \mu^+(X)}{2}=Q, \label{eqn:opt1}  
\end{equation}
Consequently,
\begin{align}
D_G(X, X') = Q - \mu^*(X) &= \frac{\mu^+(X) - \mu^*(X)}{2}\\
D_A(X, X') = Q - \mu^+(X) &= \frac{\mu^*(X) - \mu^+(X)}{2}=-D_G(X, X').
\label{eqn:opt}  
\end{align}
Thus, $D_A=-D_G$ ensures a maximal paradoxical difference.
\end{theorem}
The proof consists of two parts: a lemma on the values of the optimization criterion, followed by the actual optimal solution claim.

\begin{lemma}\label{le1}
Let $\forall x, y, k \in \mathbb{R} \setminus \lbrace - \infty, + \infty \rbrace:  ~~ y < 0 < x, k > 0 \land x + (-y) = k$, we propose that:
\begin{equation}
    \argmax_{x, y} ~~ x*(-y) = \lbrace (x, y) ~~~~ \vert ~~~ (x = -y) \land (x,y) =  \argmax_{x,y}~~ \text{abs}(x * y)\rbrace. \label{eqn:p1}
\end{equation} 
Eqn.~\ref{eqn:p1} is a generalized representation of Eqn.~\ref{eqnmax} where $x=D_G(X, X')$, $y=D_A(X, X')$, and $k = ID(X)$.
\end{lemma}

\begin{proof}[Proof of Lemma \ref{le1}]
Using an argument by contradiction, suppose that Eqn.~\ref{eqn:p1} is false, and the assumed solution of $x=-y$ is not optimal, then $\forall x, y, x' \in \mathbb{R}\setminus \lbrace \infty \rbrace ~~ \text{with}~~ k = x - y, ~ \exists ~y' \in \mathbb{R}:$
\begin{align}
    & y' < y = -x <  0 < x < -y' \nonumber\\ 
    & \land  ( x'*(-y') > x*(-y) ) 
    \nonumber\\ & \land ~~ k=x'-y'
    \label{eqn:p1false}
\end{align}
This implies $\exists~\epsilon > 0$ such that: $y'=y-\epsilon \land x'=x-\epsilon$.
For brevity $ab=a \times b$. 
\begin{align}
    (-y')x' &> (-y)x \nonumber\\
    (y-\epsilon)(x-\epsilon) &< yx \nonumber\\
    yx-\epsilon x -\epsilon y + \epsilon^2 &< yx \nonumber\\
    - \epsilon x -\epsilon y + \epsilon^2 &< 0 \nonumber\\
    - x + \epsilon &< y \nonumber\\
     x - \epsilon &> -y
\end{align}
This is a contradiction with the given $x=-y$.
As a direct consequence, the criterion in Eqn.\ref{eqnmax} then has a minimal solution when $x=D_G(X, X')=-D_A(X, X')=y$.
\end{proof}
\begin{proof}[Proof of Theorem~\ref{th:max} ($Q$ is optimal value)]
Given that $Q$ is the midpoint of the interval $[\mu^*(X), \mu*(X)]$, $Q$ is the only value where $D_G(X, X')=-D_A(X, X')$ and neither can be increased in absolute value, without decreasing the other.
Suppose $\exists ~J \neq X'$ with optimal solution $ Q'=D_G(X, J)=-D_A(X, J) $ for Eqn.\ref{eqnmax}: $Q' > Q$. We then have that $D_G(X, X')<D_G(X, J)$. In other words, suppose we can increase the geometric mean:
\begin{equation}
    \mu^*(X) < Q=\mu^*(X') < Q'=\mu^*(J).
\end{equation}
The AM-GM inequality~(Eqn.~\ref{eqnamgm}) and the stated optimality of $J$ and $Q'$ for Eqn.\ref{eqnmax}, then imply that:
\begin{equation}
      \mu^*(X) < \mu^*(X') = \mu^+(X') = Q < \mu^*(J) = \mu^+(J) = Q' < \mu^+(X).
\end{equation}
Therefore
\begin{equation}
    \vert D_A(X, J) \vert < \vert D_A(X, X') \vert,
\end{equation}
but now
\begin{equation}
    \vert D_G(X, J) \vert != \vert D_A(X, J) \vert.
\end{equation}
This is a contradiction, $J$ and $Q'$ are no longer optimal, proving that Q is optimal. 
A symmetric argument for $Q'<Q$ is obtained along the same lines.
A trivial example that $X'$ satisfying these conditions exists is $X'=[Q, Q, Q, ...]$.
\end{proof}

\begin{figure}
  \centering
  \includegraphics[keepaspectratio, width=.6\linewidth]{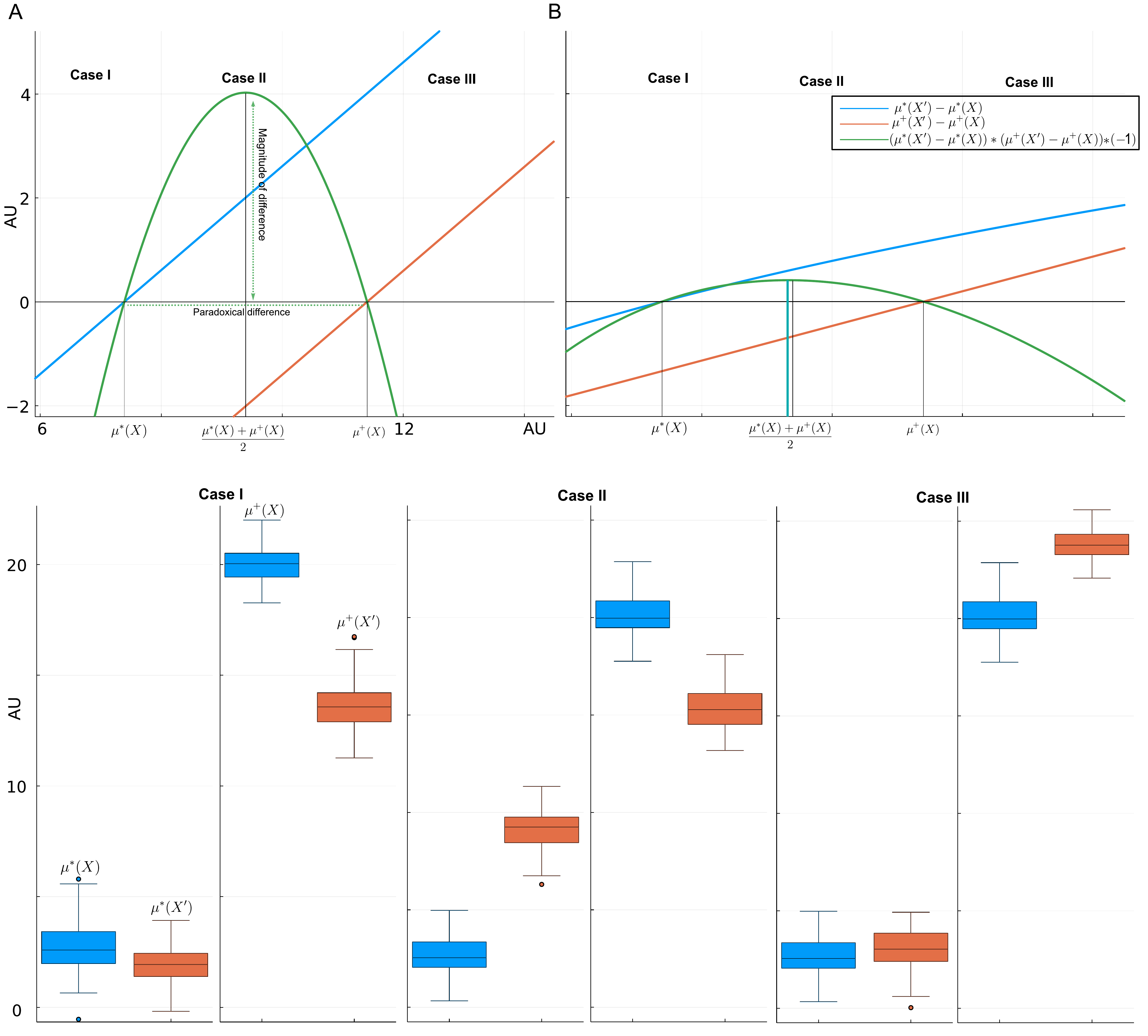}
  \caption{Given 2 vectors ($X, X'$), when their means are compared, with and without a transformation, 3 cases can occur. \internalreview{$X$ and $X'$ differ by elements on the X-axis. Cases I-III are the intervals $(0, \mu^*(X)]$, $(\mu^*(X) \mu^+(X))$, and $[\mu^+(X), +\infty)$.} In cases I and III, the means of $X$ and $X'$ are either both small (I), or large (III). However, a paradoxical difference is possible (case II), where the ordering is conditional on the transformation (log). The green curve in panels A-B, the product of both mean gradients, is negative when this paradoxical inversion occurs. We illustrate that if we have total freedom in choosing $X'$, paradoxical inversion (Eqn.~\ref{eqn:paradox}) is maximal for $\frac{\mu^*(X)+\mu^+(X)}{2}$ \internalreview{(A)}. However, when $X'= X \vectorconc [x]$, this optimal value is only approximated, not reached \internalreview{(B)}. Values are plotted on an arbitrary unit (AU) scale.}
  \label{fig:argmax}
\end{figure}

\paragraph{Approximation}
$Q$ is an optimal solution if $X$ and $X'$ can differ in any number of elements.
However, if only a few, or one element can differ, then $Q$ is only a close approximation, as the counterexample Figure~\ref{fig:argmax} illustrates.
In the left panel (A), for $X=[\exp(1), \exp(3)]$, we have total freedom to pick $X'$, and easily converge to the optimal at $Q$.
However, in the right panel (B) with $X=[\exp(1), \exp(3)]$, but $X'= X \vectorconc [x] = [\exp(1), x, \exp(3)]$, $Q$ is close but not optimal. The actual optimal value is indicated by the teal line. The constraint of only adding 1 element ensures that the gradients $\triangle_x \mu^*(X)$ (blue) and $\triangle_x \mu^+(X)$ (orange) are limited more by existing elements.
Figure~\ref{fig:argmax}\internalreview{-A, B} graphically illustrates that the paradoxical difference is constrained by the interval $[\mu^*(X) , \mu^+(X)]$.

If one wishes to express the difference in logarithmic or exponential scale, different optimality points are likely to result.
With the above conditions, we can leverage the finite difference operators to derive analytical expressions that define how one finds $X'$.
We derive this for the \replacement{} difference, the common case, given that preferentially samples of equal size are compared.
An immediate elegant consequence is the differential equation for the inter-mean distance ($ID$):
\begin{align}
    \delta_{Y,Z} ID(X) = ID[X] - ID[(X \vectorconc Y) \vectordiff Z] = ID[X] - 0 = ID[X]\label{eqn:IDX}.
\end{align}
The expressions for the \concatenation{} and \deletion{} difference can be derived along the same lines.
\subsubsection{Paradoxical replacement difference}
Combining results from Eqn~\ref{eqn:opt1} and Table~\ref{tab:opposite} we obtain:
\begin{align}
    \delta_{Y,Z} \mu^+(X) & = \frac{\mu^*(X)-\mu^+(X)}{2} \nonumber\\ & = \delta_{Y,Z} \mu^*(X).
\end{align}
This simplifies to 
\begin{align}
    \delta_{Y,Z} \mu^+(X) & = \frac{\mu^*(X)-\mu^+(X)}{2} \nonumber\\ 
    \frac{M}{N}(\mu^+(Y) -  \mu^+(Z)) & = \frac{\mu^*(X)-\mu^+(X)}{2} \nonumber\\
    \mu^+(Y) -  \mu^+(Z) & = \frac{N(\mu^*(X)-\mu^+(X))}{2M} \label{eqn:arimrep}
\end{align}
for the arithmetic mean, and
\begin{align}
    \delta_{Y,Z} \mu^*(X) & = \frac{\mu^+(X)-\mu^*(X)}{2} \nonumber\\ 
    \mu^*(X)(\frac{\mu^*(Y)}{\mu^*(Z)}^\frac{M}{N}-1) & = \frac{\mu^+(X)-\mu^*(X)}{2}\nonumber \\
    2(\frac{\mu^*(Y)}{\mu^*(Z)}^\frac{M}{N}-1) & = \frac{\mu^+(X)}{\mu^*(X)} - 1\nonumber\\
    \frac{\mu^*(Y)}{\mu^*(Z)} & = (\frac{\mu^+(X)}{2\mu^*(X)} + 1/2)^{\frac{N}{M}} \label{eqn:geomrep}
\end{align}
for the geometric mean, with $\vert Y \vert = M = \vert Z \vert$.
Eqn.~\ref{eqn:geomrep} illustrates its non-linear character, matching Fig.~\ref{fig:argmax}-B, contrasted with the linear behavior of Eqn.~\ref{eqn:arimrep}. The reciprocal of the `flatness'~\cite{rao2010adaptive} measure ($\mu^*(X)/\mu^+(X)$) is also present in Eqn.~\ref{eqn:geomrep}, illustrating its informative value.

\subsubsection{A heuristic that maximizes the paradoxical divergence}\label{subsec:heur}
\begin{figure}
  \centering
  \includegraphics[width=0.5\linewidth, keepaspectratio]{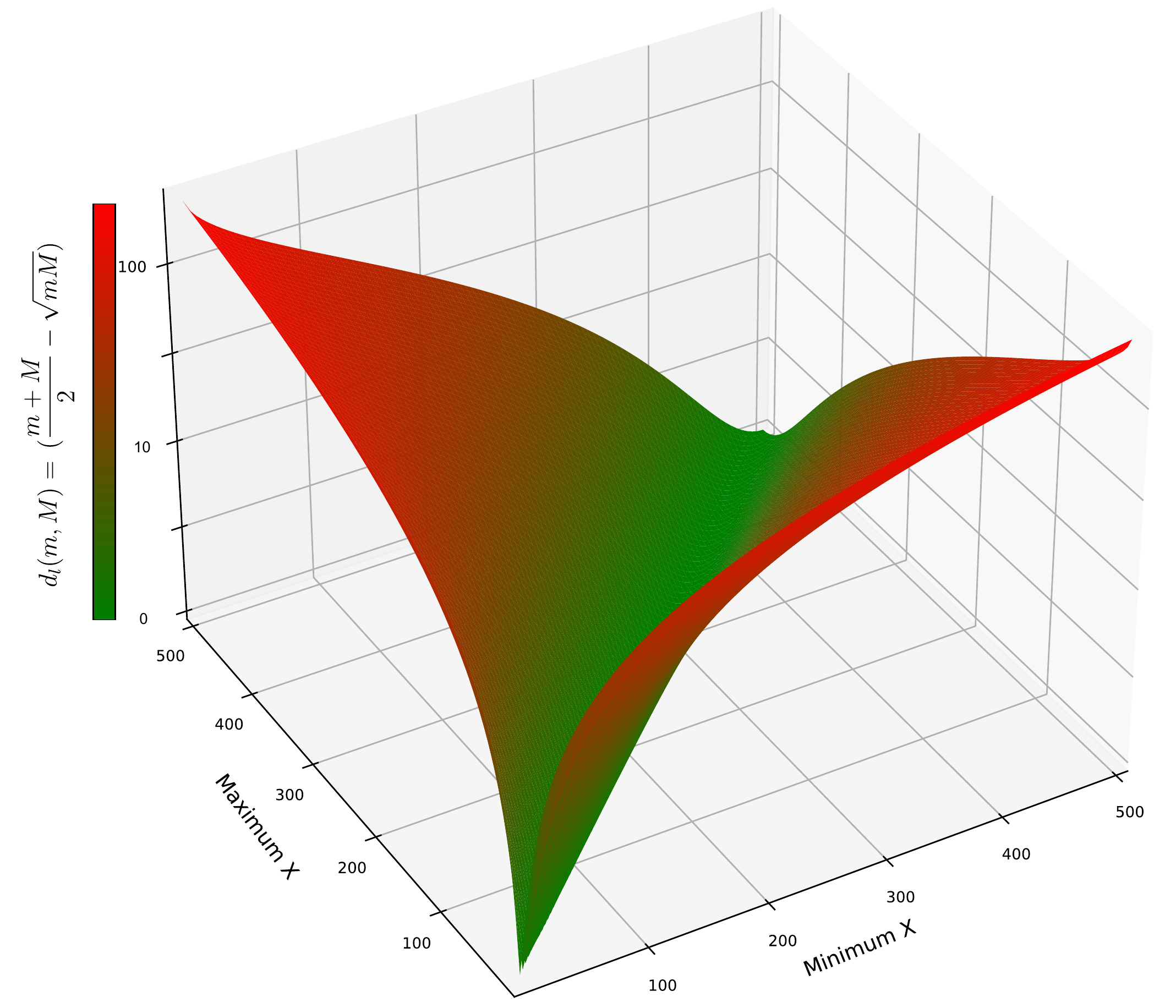}
  \caption{Illustration of how changes in minima ($m$) and maxima ($M$) of a \set{} $X$ can lead to a rapid increase in $d(m,M)$ (Z-axis, color bar, $\log_{10}$ scale), which in turn makes paradoxical comparison (Eqn.~\ref{eqn:paradox}) more likely.}
  \label{fig:vampiremesh}
\end{figure}

The above equations allow one to solve for the exact optimal transformations, if they exist, that obtain the maximal paradoxical $X'$. 
However, there may be multiple such transformations, or the exact transformation may in fact not exist. 
We can define heuristics that still approximate the result, and those heuristics can, as we shall show, give insight into real data patterns that induce the paradox.
Note first that adding elements $Y$ to $X$ such that $Y$ is chosen from the mean interval $(\mu^*(X), \mu^+(X))$ will, as a direct consequence of Eqn.~\ref{eqn:fwdid},\ref{eqn:fwdam},\ref{eqn:fwdgm}, result in an increase in the geometric mean, and a decrease in the arithmetic mean.
As a result, $ID[X \vectorconc Y] < ID[X]$.
Conversely, removing elements from $X$ outside the mean interval can increase $ID[X']$, with the largest change induced by $\text{minimum}(X)$ and \text{maximum(X)}.
This is a consequence of Eqn.~\ref{eqn:backgm} and~\ref{eqn:backam}, with $\vert Y \vert = 1$:
\begin{align}
    \argmin_{[x]} \nabla_{[x]} \mu+(X) &= \min(X) \\
    \argmin_{[x]} \nabla_{[x]} \mu*(X) &= \min(X) \label{eqn:removegmmin} \\
     \argmax_{[x]} \nabla_{[x]} \mu*(X) &= \max(X) \\
     \argmax_{[x]} \nabla_{[x]} \mu+(X) &= \max(X) 
\end{align} 
The proof naturally follows from the trivial observation that both the geometric and arithmetic mean of $[x]$ are equal to $\min([x])$ and $\max([x])$. 
Given this, suppose now that $\exists ~x' \neq \min(X)$ that is a minimal solution to Eqn.~\ref{eqn:removegmmin}.
In order to satisfy Eqn.~\ref{eqn:backgm} $x'$ must be such that $\mu^*([x']) < \mu^*([\min(X)])$, this implies that $x' < \min(X)$, which cannot be true, proving Eqn.\ref{eqn:removegmmin}. 
The other statements are proven with a symmetric argument.
Let us define our heuristic for replacements as adding $Y = [q, q]$, to replace $Z=[m, M]$, where $q=\frac{\mu^+(X)+\mu^*(X)}{2}, m=\min(X), M=\max(X)$. To keep the number of elements the same under replacement, we use $q$ twice.
Recall that from Eqn.~\ref{eqn:amrep}, \ref{eqn:gmrep} a paradoxical difference will be induced by element replacement in $X$ by $Y, Z$ if
\begin{align}
    \delta_{Y,Z} \mu^*(X) & > 0  \nonumber\\
    \frac{\mu^*(Y)}{\mu^*(Z)} & > 1 \nonumber\\
    \mu^*(Y) & > \mu^*(Z)
\end{align}
and
\begin{align}
    \delta_{Y,Z} \mu^+(X) & < 0  \nonumber\\
    \mu^+(Y) & < \mu^+(Z).  
\end{align}
The geometric and arithmetic mean of $Z=[m, M]$ are $\sqrt{m*M}$ and $\frac{m+M}{2}$ respectively, and so we get
\begin{equation}
    \sqrt{m * M} < \frac{\mu^*(X)+\mu^+(X)}{2} < \frac{m + M}{2}.\label{eqn:heuristic}
\end{equation}
Equation~\ref{eqn:heuristic} determines when the heuristic will induce a paradoxical difference.
Let $Q=\frac{\mu^*(X)+\mu^+(X)}{2}$, substituting the results from Table~\ref{tab:sumdiff}, the heuristic will induce a maximal paradoxical difference if:
\begin{align}
    \delta_{[Q,Q],[m, M]} \mu^+(X) & = Q - \mu^+(X)  \nonumber\\
    \frac{2}{N}(Q-\frac{m+M}{2})& = Q - \mu^+(X)
\end{align}
and
\begin{align}
    \delta_{[Q,Q],[m, M]} \mu^*(X) & = Q - \mu^*(X) \nonumber \nonumber\\
    \mu^*(X)(\frac{Q}{\sqrt{m*M}}^\frac{2}{N}-1)& = Q - \mu^*(X)\nonumber\\
    (\frac{Q}{\sqrt{m*M}}^\frac{2}{N}-1)& = \frac{Q}{\mu^*(X)} - 1.
\end{align}
For any set of real numbers $\vert X \vert \geq 1$, $m$ and $M$ always exist, as does $Q$, which we insert twice to keep the size of $X'$ identical to that of $X$.
To evaluate the minimal conditions for the heuristic (Eqn.\ref{eqn:heuristic}), let us now define:
\begin{equation}
    d(m, M) = \frac{m+M}{2}-\sqrt{m*M}. \label{eqn:dv}
\end{equation}
We note, as illustrated in Fig.~\ref{fig:vampiremesh}, that a larger value of $d(m,M)$, by decreasing $m$ or increasing $M$, corresponds with a higher likelihood of inducing paradoxical comparison (Eqn.~\ref{eqn:paradox}).
This is important for practitioners, because decreasing $m$ or increasing $M$ can occur when one removes outliers, when extreme values are present, when data is resampled, or even when segmentation of objects fuses or splits objects incorrectly, leading to unexpected stark changes in the mean and risking introducing the paradox.
While long tail distributions are often asymmetric, the results in Fig.~\ref{fig:vampiremesh} should convince the practitioner that even in symmetric distributions the paradox can appear.
However, the paradox can also be a consequence of subtle shifts in the data, corresponding with valuable domain specific findings, as we explore in Section~\ref{sec:results}.

\section{Empirical results}\label{sec:results}
We first start with a handcrafted example using an exponential distribution that mimics real world data in Section~\ref{sec:emp}. 
Given that resampling is frequently employed in quantification of image data, we next review how the paradox behaves under resampling in Section~\ref{sec:resample}.
Finally, we end with a cellular biology inspired use case, illustrating that even small frequency shifts in protein structures, modelled by a Markov chain, can induce the paradox.
\subsection{Prevalence of paradox on empirical distribution}\label{sec:emp}
\begin{figure}[htbp]
  \centering
  \includegraphics[keepaspectratio, width=0.70\linewidth]{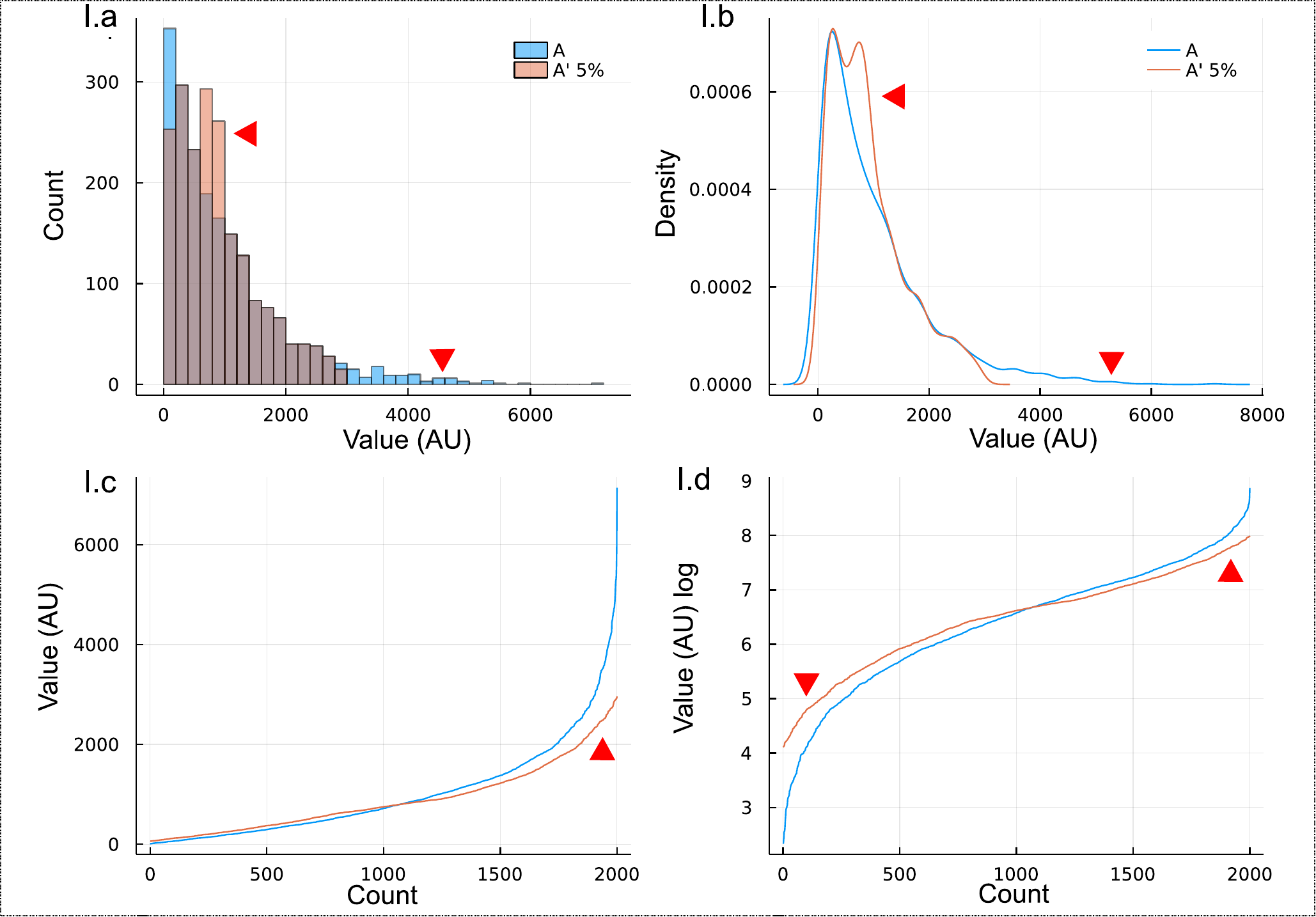}
  \caption{(I) Long tail data distributions ($A, A'$) used in experiments (I.a-d).
  Note: To avoid confusion with the X-axis, in the figure legend we use the notations $A$ and $A'$, instead of $X$ and $X'$, to refer to \set{}s.
  Randomly replacing 5\% of $A$ already dramatically shortens the tail, both in histogram (I.a) and density plot (I.b), even though only a minor fraction of $A$ is altered. 
  Note how in the plots of the sorted values (I.c-d) the log-transform (I.d) responds near symmetrically to the change, but in the non-transformed space (I.c) this is not the case, highlighting why the arithmetic and geometric means behave differently to such changes.}
  \label{fig:adist}
\end{figure}

\begin{figure}[htbp]
  \centering
  \includegraphics[keepaspectratio, width=0.650\linewidth]{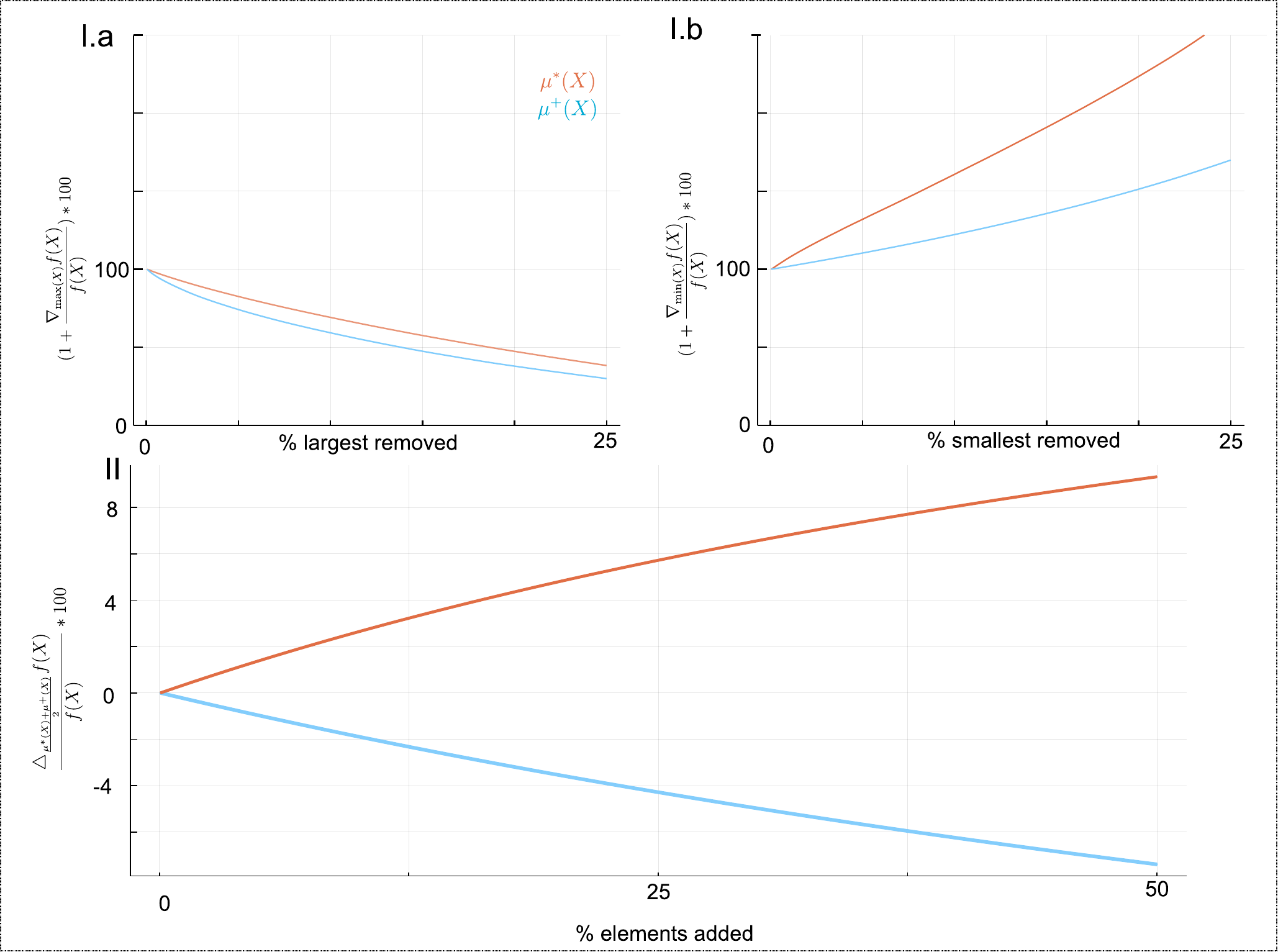}
  \caption{
  (I) Removing the $k^{th}$ largest (I.a) and smallest (I.b) values (without replacement) illustrates how both means respond differently and how the inter-mean ratio and interval behave. 
  Note that removing small values is numerically equivalent to adding or re-weighting the largest values. 
  The stronger change observed in removing the smallest values underlines, therefore, the large contribution extreme values have on the means.
  (II) Behavior of mean divergence under element insertion (Alg.~\ref{alg:insert}). At each step $x$ (X-axis) a new element $Q=\frac{\mu^*_{Ax}+\mu^+_{Ax}}{2}$ is added, and 
  $\delta^*$, $\delta^+$ computed (Y-axis). For a derivation of $Q$ see Section~\ref{subsec:heur}.
  $A$ contains at the start 2000 elements of an exponential distribution. 
  The divergence of means increases markedly.}
  \label{fig:ain}
\end{figure}

\begin{figure}[htbp]
  \centering
  \includegraphics[keepaspectratio, width=0.75\linewidth]{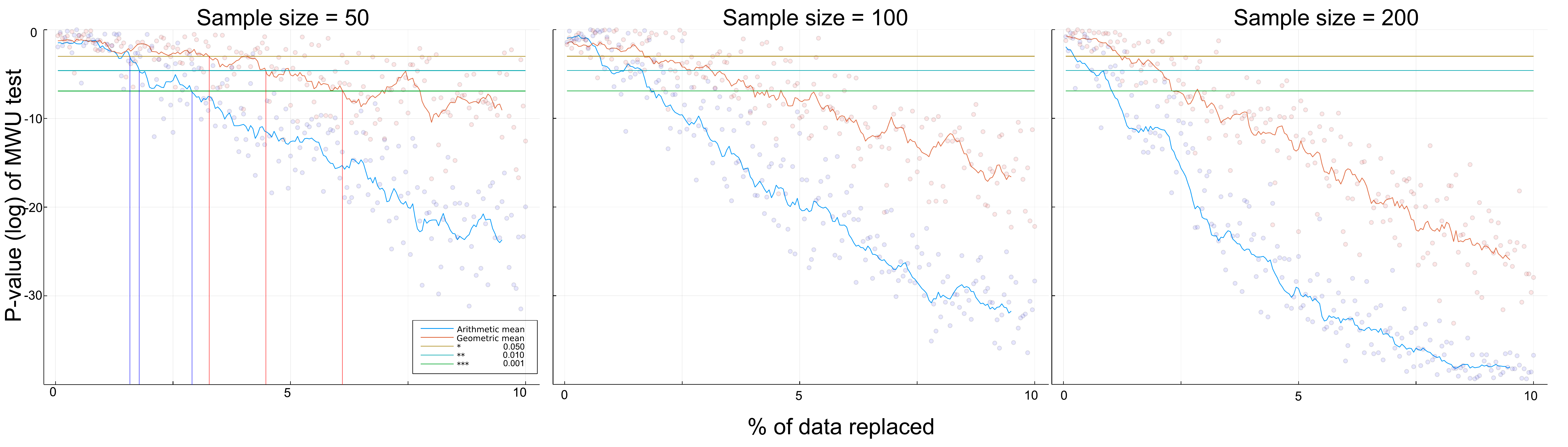}
  \caption{The prevalence and consistency of the log-paradox under aggregation and resampling. 
  The term consistency refers to convergence to a known pattern as a parameter (x) increases. 
  We report the p-value of the Mann Whitney U-test, testing if the aggregated means are different (Y-axis, log scale). 
  Vertical lines plot the first point where significance for one of the three commonly used thresholds (horizontal lines, 0.05-*, 0.01-**, 0.001-***) is exceeded.
  Resampling is done 50 ($N$) times, with sample size ($S$) 50, 100, and 200. Up to 200 elements (10\%) of $A$ are replaced to form $A'$. \internalreview{The replaced value is chosen randomly from the range $[\mu^*(A), \mu^+(A)]$}. The three standard levels of $\alpha$ thresholds of significance are plotted horizontally. A smoothed (average with window size 10) line is plotted between the data points. The faster decrease of p-values for larger sample size is not due to the actual sample size, larger resampling from the source data ($A$) means that extreme values, more likely captured, are maximizing divergence. The log-paradox is, therefore, consistently present under aggregation and resampling. 
  }
  \label{fig:repace}
\end{figure}

We now set out to answer the practitioner's question: "On realistic data, how quickly does the paradox appear, and how large can its significance get?"
We create a simple exponentially distributed \set{} of numbers $A$, sampled from a scaled probability density function (pdf) 
\begin{equation}
    f(x) = 10 + e^{-x} * 1000. \label{eqn:gena}
\end{equation}
The distribution of $f$, comprising 2000 samples, is described in Figure~\ref{fig:adist}-I.a-b.
The offset of 10 and scaling of 1000 ensures no untoward effects (Section~\ref{subsec:choiceb}) are induced by logarithm base 10, and reflect the long tail nature of real world exponential distributions.
Sizes of objects recorded in biomedical 2D and 3D images are often distributed with a long tail, due to the quadratic (2D) or cubic (3D) increase of surface/volume. 
Reasons for a long tail are diverse: the objects can be a mixture of classes (e.g. different cell types, lesions, organs), or a mixture of structures that aggregate from many small into a few large (e.g. protein subcellular structures), segmentation can fail to separate proximate objects, doubling their size, or other domain specific reasons. 
For example, research involving characterizing pathological conditions, such as aggregation of plaques, for example amyloid-$\beta$ in Alzheimer disease~\cite{hart2016ocular}, or uncontrolled growth as manifest in malignant cancers, will likely encounter a low frequency of very large objects in the presence of high frequency of small objects. 
All of these use cases can lead to a low frequency of very large objects, leading to a classic long tail distribution for object size measurements. 
However, a long tail is not necessarily solely at the larger end of the distribution; the tail can extend to the minima, or both sides. 
A low frequency of very small objects can be present with a higher frequency of medium-sized objects, for instance, when noise (presenting as very small objects) is not completely filtered out.
As we will see later (Section~\ref{exp:iomed}), reporting object sizes in an emergent system, where smaller objects aggregate to form larger objects, is a prime example of a situation where a long tail distribution, and subsequent log paradox, can arise. 
In non-biomedical settings, wealth distributions (Pareto) would follow a similar long tail distribution.
Next, we test if we can induce the log-paradox in $A$ by applying Alg.~\ref{alg:insert} to illustrate $\triangle_Y f(X)$ difference (Eqn.\ref{eqn:concdif}), and Alg.~\ref{alg:replace} for the \replacement{} difference $\delta_{Y,Z} f(X)$ (Eqn.~\ref{eqn:repdif}).
\internalreview{Figure~\ref{fig:ain} shows the results when we remove values (I.a-b) or insert $Q$ (II), with insertion clearly inducing the paradox.}
Because the arithmetic mean is very sensitive to extreme values, we can fairly easily induce the log paradox both under the insertion and replacement scenarios by using the heuristics from Eqn.\ref{eqn:heuristic}. 
Even under random replacement (Fig.~\ref{fig:repace}-A) the divergence is consistent, and a marked reversal is obtained even at 10\% (200 out of 2000) replacements.
\internalreview{Observe that e}ven 5\% is sufficient to cause marked changes in distribution shape.
The replacement of minimum and maximum, respectively, (as shown in Fig.~\ref{fig:ain}-I.a-b) illustrates the sensitivity of the arithmetic mean to extreme values. The difference in sensitivity to extreme values between the two means drives the divergence (ID(X)).



\subsection{Paradox in presence of aggregation and resampling}\label{sec:resample}
To test if the paradox is present under resampling, we take our generated data sequence $A$ (Eqn.\ref{eqn:gena}), and repeatedly sample with replacement, as per the bootstrap protocol (Alg.~\ref{alg:sample}). We test 3 sample sizes: 50, 100, and 200.
We compare resampling on $A'$ where we replace up to 10\% of $A$ with points sampled from the interval [$\mu^*(A), \mu^+(A)$] (Alg.~\ref{alg:replace}).
Results are shown in Fig~\ref{fig:repace}).
For each pair of samples ($B_i, B'_i$) the geometric and arithmetic mean are computed, resulting in 2 distributions of sample means.
We test the hypothesis that they differ using the Mann Whitney's U-test (MWU), a commonly used non-parametric test, and report the p-values.
We only report if the difference is paradoxical, where the geometric mean \textbf{increases} while the arithmetic \textbf{decreases}.
Note that the choice of resampling or aggregation practice as well as the best test, \eg MWU, would depend on the research question and the underlying data distribution. 
Here, we are more interested in observing if the log-paradox is present under these reporting conditions, and if it is, how does the pattern behave under different choices of parameters.
The size of the sampled data points is set to vary between 50-200 to balance between statistical power and inducing significance by data size alone.
Note how the increase of the sample size of $A$ and $A'$ ($S$) rapidly leads to `significance'.
Increasing the sample size results in a higher probability that extreme values are included.
In turn, the extreme values can make the paradox more likely to appear.
With only 5\% of $A$ altered by random replacements (Fig.~\ref{fig:repace} X=100), it is already trivial to get p values $<$g 0.001, validating that even with random replacement the paradox is easily and consistently induced. 
A smoothed line is plotted as a moving (window=10) average over each data point.
Confirming earlier results and prior knowledge, the arithmetic mean (blue) is markedly easier to change from its initial value.
From this experiment, it is clear that resampling does not necessarily prevent the paradox.

\subsection{Biomedical imaging \internalreview{inspired} use case}\label{exp:iomed}
\begin{figure}[htbp]
  \centering
  \includegraphics[width=0.65\linewidth, keepaspectratio]{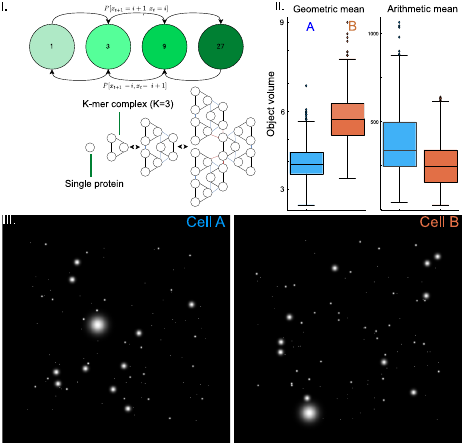}
  \caption{Aggregation - destruction dynamics of protein complexes in a cell modelled by a Markov chain. Single proteins combine into k-mers, determined by environmental conditions. Here we show a simple 3-way nested aggregation of 3-mers. The Markov chain has a directional arc between nodes $i$ and $j$ weighted by the probability $P[x_{t+1}=j|x_{t}=i]$ determining the likelihood that a complex of type $i$ at time $t$ will transform into a complex of type $j$ at time $t+1$.
  II: We sample 1000 virtual cells from the probability distribution in Table~\ref{tab:markov} ({\color{blue}{A}}, {\color{orange}{B}}), and measure the mean volume of complexes per cell. The box plots show how the geometric means show an inverse group difference compared to the arithmetic means. III: Two images sampled from the Markov chain models in Table~\ref{tab:markov}. The images illustrate that the difference in the protein size distribution may not be apparent visually, yet it can lead to markedly different quantification. Furthermore, conditional on log-transform, the analyst can report paradoxical conclusions. 
  }
  \label{fig:markov}
\end{figure}

\internalreview{In this use case, we model with a Markov chain, as illustrated in Figure~\ref{fig:markov}-I, the construction/destruction dynamics of protein complexes and the subsequent quantification of images that capture such structures.}
A protein structure, or a complex, is composed of one or more k-mers, where $k$ indicates the number of individual proteins forming the structure. 
Physics and the biological environment determine how multiple proteins aggregate into a k-mer, and how multiple k-mers aggregate to form a larger protein structure.
A directed arc from node $i$ to node $j$ is weighted by the probability, $P[x_{t+1}=i|x_{t}=j]$, that a complex $x$ of type $i$ at time $t$ will transform to type $j$ at time $t+1$. 
For example, if the arc 1 $\rightarrow$ 3 is weighted 0.3, the probability of a single protein forming a trimer would be 30\%.
We present a simple in silico experiment using two different example cell lines A and B, where proteins combine in nested trimers (Fig.~\ref{fig:markov}-I), the distribution is subtly different with B favoring trimers over 27-mers and single proteins.
In Table~\ref{tab:markov} we list the number of structures per cell, normalized to an identical amount of proteins, and the frequency distribution.
We next record the volume of these structures, estimated by $k^{3}$.
Note that very large structures, even though they make up a tiny fraction of the probability distribution (Table~\ref{tab:markov}), will appear visually striking due to their size, as observed in Fig.~\ref{fig:markov}-III.
From each distribution, we sample 1000 `cells' consisting of 525 k-mers each, simulating a large controlled experiment.
Example images (2D) of 1 cell of each type are visualized in Fig.~\ref{fig:markov}-III.
In Figure~\ref{fig:markov}-II, we plot the distribution of geometric and arithmetic means of each `cell' as an estimate of the population mean. 
The research question "is the volume of protein structures in A $>$ B?" now has two contradicting answers, because $\mathbb{E}[\log.(A)] < \mathbb{E}[\log.(B)]$ while $\mathbb{E}[A] > \mathbb{E}[B]$. 
This in silico example illustrates that the presence of the log-paradox in fact indicates complex underlying dynamics that differentiate cell lines A and B.

\section{Discussion}\label{sec:disc}
The presence of the log paradox can lead practitioners to favor abandoning the log-transform altogether, similar to what Feng et al.~\cite{changyong2014log} argue, in favor of models such as generalized linear models~\cite{liang1986longitudinal}. 
However, we argue that the paradox is informative and not a reason to discard the log-transform. 
We identify the necessary and sufficient conditions required for the paradox to appear.
Armed with this information, rather than discarding the transformed results, the practitioners can now base their selection and application of more complex discovery methods on these preconditions.
We have illustrated that a hidden Markov chain model can induce the paradox (Fig.~\ref{fig:markov}), and refer the interested reader to Cowles et al.~\cite{cowles1996markov} for an in depth review. 
Markov chains are one example of recovering a mathematical model for the abstracted `generating function'.
The Markov chain abstraction and use case of Section~\ref{exp:iomed} is modelled on dynamics in subcellular biology.

For example, mitochondria, large subcellular organelles, respond to conditions of starvation, stress, and energy requirements by fusion (combining) and fission (splitting).
The dysfunction of mitochondrial dynamics is increasingly linked to a host of degenerative diseases~\cite{aging, mitodegen}. Another example, caveolae, which are $\sim$100 nanometer (nm) diameter spherical invaginations of the cell membrane that are responsible for, among others, enabling the cell membrane to adapt to high physiological stressors, have similar (dis)aggregation dynamics~\cite{bastiani2010caveolae, STAN2005334, KURZCHALIA1999424} and large impact on organism health~\cite{cohencav}. 
A cell continuously constructs and decomposes caveolae from and into scaffolds in response to physical, chemical, and genetic triggers, creating caveolae from smaller ($\sim$10-20 nm) components and decomposing them into smaller \internalreview{components~\cite{khater2019super}}.
In these cases, the size of an object (\eg k-mer, mitochondrion, caveolae) is strongly inversely related to its frequency, \ie a larger protein/mitochondrion/caveolae would be less frequently observed than a smaller protein/mitochondrion/caveolae, hence leading to long tail distributions.

Recently, improved causal discovery algorithms~\cite{sldisco, iter, huang2018generalized} have been proposed, where a graph causal model that explains the data generation function can be recovered.
Interpretable models are by no means the only direction to take in response to the paradox.
For example, recent work in weak or unsupervised methods that identify contrasting subpopulations~\cite{peng2022generalized} demonstrates a different approach to the same problem statement.
If the underlying problem is one based on quantifying extreme values, then leveraging extreme value theory inspired models is appropriate~\cite{de2006extreme, smith1990extreme}.
If anything, the paradox is a lead that allows practitioners to dig deeper into the data and reveal the underlying patterns that induce the paradox.

While we have so far discussed right long tails, where extreme values are greater than the mean of a distribution and thus located on the right on a numerical axis, none of the conditions we compute restrict the paradox to the right long tail only.
As the derivation of the heuristic (Section~\ref{subsec:heur}) shows, there is no reason why the paradox cannot occur on the left long tail distributions. 
Symmetry does not preclude the paradox either.
Recall, however, that the paradox can only be induced if the inter-mean distance is sufficiently large, and this is more likely to occur for distributions where the majority of values is far larger than the base of the logarithm, yet with sufficiently large minima and maxima. 
We show an in-silico example in Fig.~\ref{fig:symm}.
A Normal distribution, by definition symmetric, is varied across its $\mu=x$ parameter, and a left tail and right tail are induced minimally by adding $\sqrt{x}$ and $x^2$, respectively.
We plot $ID(A)$ and observe that as $\mu$ increases, the values of the tails change and, consequently, $ID(A)$ increases, resulting in a larger sensitivity to the paradox.
In this paper we focused by starting from a \set{} $X$, with a large ID(X), then find $Y$, such that we can find a paradoxical comparison (Eqn.~\ref{eqn:paradox}) between them, invariably by decreasing $ID(Y)$.
However, a symmetric argument holds, given $ID(X)=0$ you can induce a paradoxical comparison with $Y$ by increasing $ID(Y)$.

The paradox can serve as a safeguard or `canary' in pre- and post-processing as well.
We documented how even simple segmentation differences can induce the paradox.
The paradox can then serve as a test for the robustness of segmentation results across methods or parameter values.
The selection of base in the log-transform can be of help here as well, as it determines the sensitivity of the transform with respect to values bounded by the base.
While much of the focus has been on data sampled from long tail distributions, that are often asymmetric, the results of the heuristic (Section~\ref{subsec:heur}) should make it apparent that the paradox is also possible in data sampled from symmetric distributions.
Finally, reproducibility requires that both transformed and untransformed data are always made available.
\section{Conclusion}
{\color{black}We have shown that long tail asymmetric distributions are sufficient conditions for appearance of the log-paradox and defined conditions under which its effect is maximal. We derived necessary and sufficient conditions for its occurrence, illustrating to practitioners when and why this paradox is likely to appear.}
Rather than a cautionary tale, we described that the paradox reveals underlying patterns in the data. 
We empirically \internalreview{show} how resampling does not prevent the paradox.
Using a simplified real-world use case from biomedical imaging, we illustrated that the study of dynamics of (de)construction of subcellular objects is susceptible \internalreview{to the log-paradox}, and make recommendations on how to move beyond the paradox to report deconfounded findings on the underlying data.
In general, proportionally distributed long tail data can be sufficient to induce the paradox, a reminder to properly reflect on the independence of samples in data where resource constraints can skew proportional values, such as object sizes in cells.
The log-paradox is a useful indicator that allows revealing the underlying complex pattern with more powerful, intuitive, but perhaps not always available methods, that capture the data generation model.

\bibliographystyle{imsart-number} 
\bibliography{bibliography}  

\clearpage 

\begin{appendix}

\section{Software}\label{sec5}
Software in the form of Julia code, together with a sample
input data set, and complete documentation is available under an open source license at \\ \url{https://github.com/bencardoen/LogParadox.jl}.
A Zenodo doi linking to the version matching this manuscript can be found at \\ \url{https://doi.org/10.5281/zenodo.7545842}.

\section{The importance and consequences of choice of base}\label{subsec:choiceb}
\begin{figure}[htbp]
  \centering
  \includegraphics[width=0.8\linewidth, keepaspectratio]{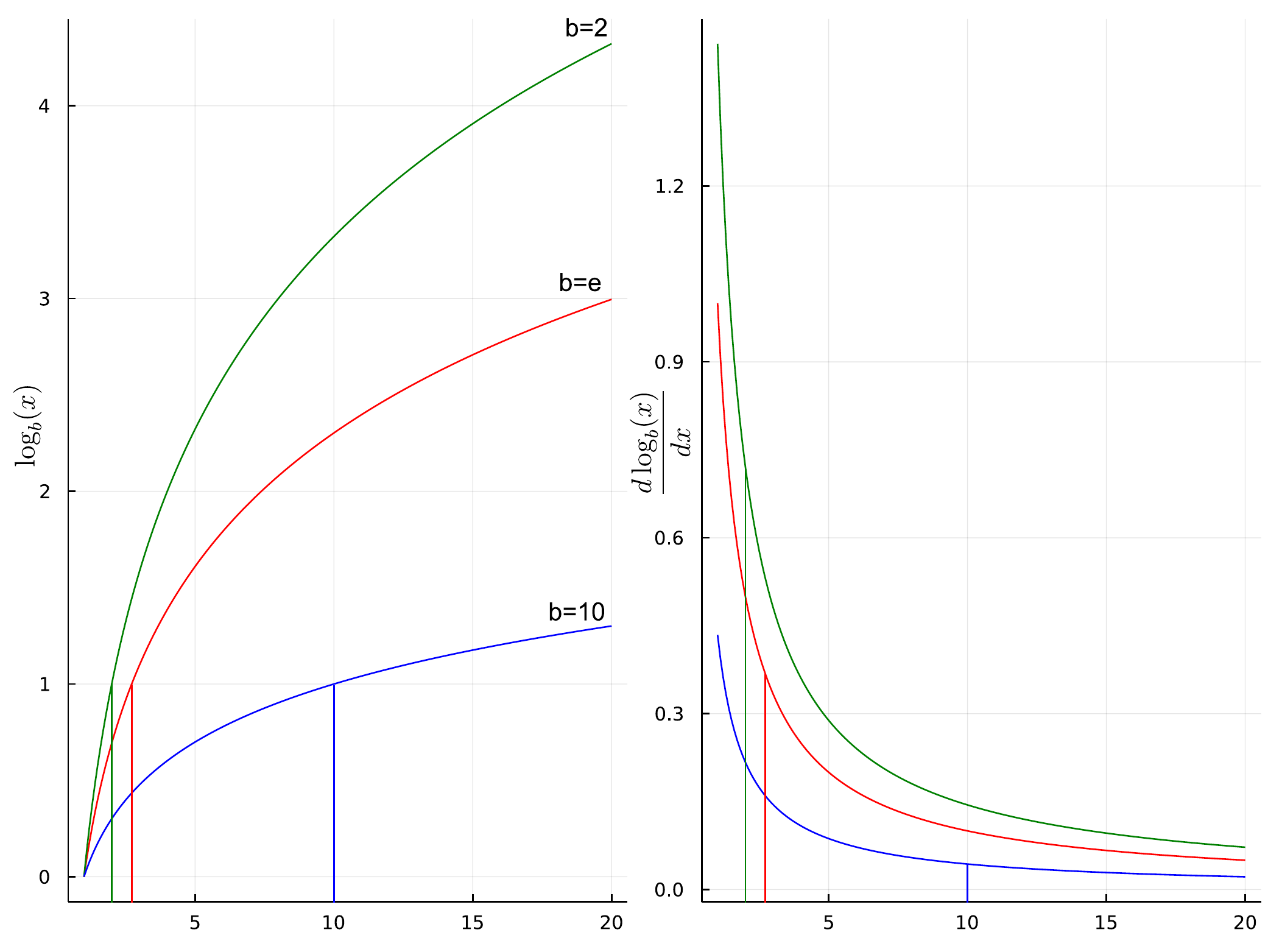}
  \caption{Illustration of the importance of choice of base and its asymmetric effects in context of distribution. The left panel shows $\log_b(x)$ for $b=[2, e, 10]$, the right the corresponding first derivative with respect to $x: \frac{d \log_b(x)}{dx} = \frac{1}{\log_e(b) x}$. Vertical lines mark the base $b$. Note how the magnitude of the derivatives is \internalreview{larger for $x<b$ and $x\rightarrow0$, compared to $x>b$}. Therefore, if a subset $A\subset X$ and $\lbrace a < b \vert \forall a \in A \rbrace$, then $A$ can bias the transformed data $\log_b.(X)$. For this reason, one can sets $b < \min(X)$ to avoid transform induced biases.}
  \label{fig:logd}
\end{figure}

Before we move to the empirical results, it is important to illustrate what effects the choice of base $b$ has on log-transformed data.
\internalreview{Figure~\ref{fig:logd} plots the logarithm and its derivative (Y-axis) for three common bases (2, $e$, 10).
It is clear to see that the derivative of the logarithm base $b$ is largest for $x < b$.}
More than that, $\lim_{x \rightarrow 0} \frac{d \log_b(x)}{dx} = + \infty $.
\internalreview{Intuitively, captured by the derivative, the log-transform changes very quickly for values $x < b$, compared to values $x > b$.}
Therefore, by ensuring that $\forall x \in X: x > b$ we avoid asymmetric effects induced by subsets of the data conditional on the choice of $b$.

\clearpage
\section{Paradox on symmetric and left tail distributions}
\begin{figure}[htbp]
  \centering
  \includegraphics[width=0.45\linewidth, keepaspectratio]{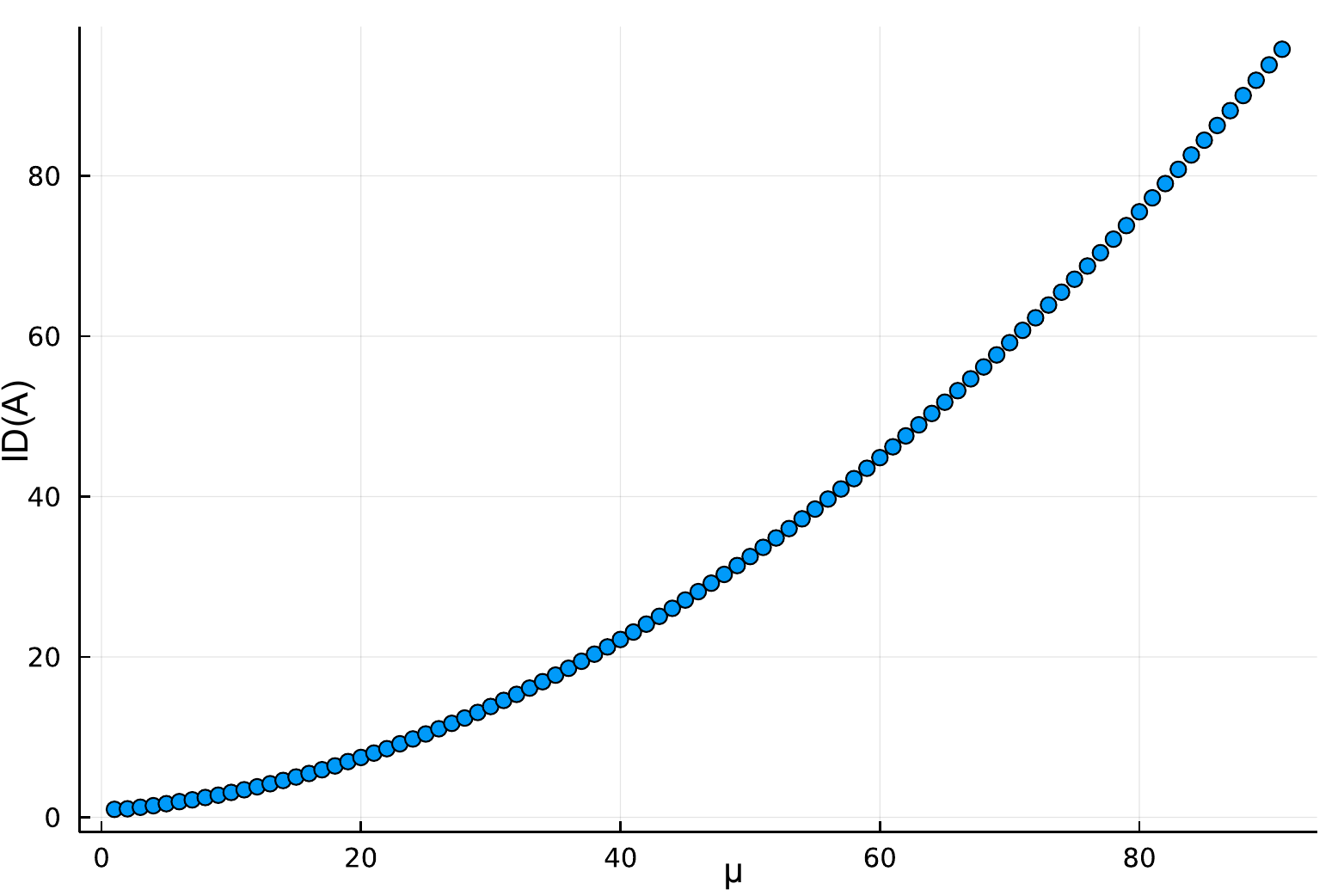}
  \caption{Sensitivity to paradox on symmetric distributions with long tails on the left and right. A \set~$A$ with 100 values, sampled from a Normal distribution $\mathrm{N}(\mu=x, \sigma=2$), is used. To simulate a long tail on both sides $\sqrt(x)$ and $x^2$ are added and $\mu$, the parameter of the distribution, is varied from 10 to 100 in increments of 1 (X-axis). For example, for $x=16$, $A=[4] \vectorconc \mathrm{N}(\mu=16, \sigma=2 \vectorconc [256]$. From the plot of inter-mean distance $ID(A)$, we observe that $ID(A)$ is easily large enough to induce a paradoxical comparison, hence, showing that symmetric and left long tail distributions are equally vulnerable.}
  \label{fig:symm}
\end{figure}

\clearpage
\section{Algorithms}
\begin{algorithm}
\caption{Induce paradox with an insert heuristic.}
\begin{algorithmic}
\small
\State \textbf{Input:} Vector $A$
\State \textbf{Output:} Vector $A'$
\State $N \gets \vert A \vert $
\State $\mu^* \gets \exp \frac{\sum \log (A)}{N}$
\State $\mu^+ \gets \frac{\sum (A)}{N}$
\State $A' \gets A \cup \frac{\mu^* + \mu^+}{2}$
\State \textbf{Return} $A'$
\end{algorithmic}
\label{alg:insert}
\end{algorithm}

\begin{algorithm}
\caption{Induce paradox with a replacement heuristic.}
\begin{algorithmic}
\State \textbf{Input:} Vector $A$
\State \textbf{Output:} Vector $A'$
\State  $N \gets \vert A \vert $
\State  $\mu^* \gets \exp \frac{\sum \log (A)}{N}$
\State  $\mu^+ \gets \frac{\sum (A)}{N}$
\State  $A = A \vectordiff \text{select}(A)$ \Comment{select={random, min, max, (min,max)}}
\State  $A' \gets A \vectorconc \text{sample}_\text{uniform}([\mu^* , \mu^+])$
\State \textbf{Return} $A'$
\end{algorithmic}
\label{alg:replace}
\end{algorithm}

\begin{algorithm}
\caption{Bootstrap Resampling}
\begin{algorithmic}
\State \textbf{Input:} Vector $A$, sample size $N$, sample frequency $M$, statistic $f$
\State \textbf{Output:} Vector $B$: sample distribution of statistic $f$
\State $B \gets []$
\For{i $\in$ $1$ .. $M$}
\State $B[i] \gets f(\text{sample}_\text{uniform}(A, N))$
\EndFor
\State \textbf{Return} $B$
\end{algorithmic}
\label{alg:sample}
\end{algorithm}
\clearpage
\section{Tables}
\begin{table}[!h]
\caption{Reference table of definitions. $X \in \mathbb{R}^N, Y \in \mathbb{R}^M, Z \in \mathbb{R}^K$, with $N, M, K \in \mathbb{N}_{>0}$, and $K < N$. Note that the expected value applies to countable finite distributions, which covers to the data discussed in this paper.}
\begin{tabular}{llll}
\hline
Name & Symbol  & Definition  & Equation \\ \hline
Vector &$X$  & $X=[x_1, .., x_n] \in \mathbb{R}^{\vert X \vert}$  & \ref{eqn:vector} \\ \hline
Expected  value & $\mathbb{E}(X)$  & $\frac{1}{\vert X \vert} \sum\limits_{x_i \in X} x_i$  & - \\ \hline
Arithmetic mean & $\mu_X^+$  & $\frac{1}{\vert X \vert} \sum\limits_{x_i \in X} x_i = \mathbb{E}(X)$   & \ref{eqn:am} \\ \hline
Geometric mean & $\mu_X^*$  & $\sqrt[\vert X \vert]{\prod\limits_{x_i \in X} x_i} = \mathbb{E}(\log.(X))$  & \ref{eqn:gm} \\ \hline
Inter-mean distance (ID) & $ID(X)$  & $\mu_X^+ - \mu_X^*$  & \ref{eqn:id} \\ \hline
Vectorized function & $f.(X), F(X)$  & $[f(x) ~\forall x \in X], f: \mathbb{R} \rightarrow \mathbb{R}$   & \ref{eqn:vfun} \\ \hline
Vector concatenation & $X \vectorconc Y$  & $[x_1, .., x_n, y_1, .., y_m]~\forall x \in X, y \in Y$  & \ref{eqn:vconc} \\ \hline
Vector difference & $X \vectordiff Y$  & $[x_1, .., x_j]~\forall x \in X \land x \notin Y$  & \ref{eqn:vdiff} \\ \hline
Paradoxical comparison & $X > Y \land f.(X) < f.(Y) $  & $\mathbb{E}[X] > \mathbb{E}[Y] \land \mathbb{E}[f.(X)] <\mathbb{E}[f.(Y)]$  & \ref{eqn:paradox} \\ \hline
Finite difference (FD) & $D_{G, Y} F(X) $  & $F(G_Y(X))-F(X)$ & \ref{eqn:fd} \\ \hline
FD under concatenation & $\triangle_Y F(X) $  & $F(X \vectorconc Y) - F(X)$  & \ref{eqn:concdif} \\ \hline
FD under deletion & $\nabla_Y F(X) $  & $F(X \vectordiff Y) - F(X)$  & \ref{eqn:deldif} \\ \hline
FD under replacement & $\delta_{Y, Z} F(X)$  & $F((X \vectorconc Y) \vectordiff Z)) - F(X)$  & \ref{eqn:repdif} \\ \hline
\end{tabular}

\label{tab:definitions}
\end{table}

\begin{table}[!h]
\centering
\caption{This table answers the question: `How does the arithmetic, geometric mean, and the mean difference for vector $X$ change given a perturbation defined by $Y$ and/or $Z$?'. The answers are given in closed form expressions, with the introducing equation linked in the rightmost column. $X, Y, Z\in \mathbb{R}^N, \mathbb{R}^M, \mathbb{R}^K, \vert X \vert = N, \vert Y \vert = M, \vert Z \vert = K$. For the \replacement{} difference, the simplified case of $\vert Y \vert = \vert Z \vert$ is listed. We refer to Table~\ref{tab:definitions} for the complete definitions of symbols used here.}
\begin{tabular}{lll}
 Finite difference & Expression  & Introducing Equation  \\[5pt] \hline
 &&\\[-1em]
 $\triangle_Y \mu^{+}(X)$ & $\frac{M}{N+M}(\mu^+(Y) - \mu^+(X))$  & \ref{eqn:fwdam1}  \\[5pt] \hline
 &&\\[-1em]
 $\triangle_Y \mu^{*}(X)$ & $\mu^*(X)(\mu^*(X)^\frac{-M}{N+M} \mu^*(Y)^\frac{M}{N+M} - 1)$  & \ref{eqn:fwdgm}  \\[5pt] \hline
 &&\\[-1em]
 $\triangle_Y \text{ID}(X)$ & $\triangle_Y \mu^+(X) - \triangle_Y \mu^*(X)$  & \ref{eqn:fwdid}  \\[5pt] \hline
 &&\\[-1em]
 $\nabla_Y \mu^{+}(X)$ & $\frac{M}{N-M}(\mu^+(X)-\mu^+(Y))$  & \ref{eqn:backam}  \\[5pt] \hline
 &&\\[-1em]
 $\nabla_Y \mu^{*}(X)$ & $\mu^*(X)^\frac{N}{N-M} \mu^*(Y)^\frac{-M}{N-M} - \mu^*(X))$  & \ref{eqn:backgm}  \\[5pt] \hline
 &&\\[-1em]
 $\nabla_Y \text{ID}(X)$ & $\nabla_Y \mu^+(X) - \nabla_Y \mu^*(X)$  & \ref{eqn:backid}  \\[5pt] \hline
 &&\\[-1em]
$\delta_{Y,Z} \mu^{+}(X)$ & $\frac{M}{N}(\mu^+(Y)-\mu^+(Z))$  & \ref{eqn:delam}  \\[5pt] \hline
&&\\[-1em]
 $\delta_{Y,Z} \mu^{*}(X)$ & $\mu^*(X)( \frac{\mu^*(Y)^{M/N}}{\mu^* (Z)^{M/N}} -  1)$  & \ref{eqn:delgm}  \\[5pt] \hline
 &&\\[-1em]
 $\delta_{Y,Z} \text{ID}(X)$ &$\delta_{Y,Z}\mu^+(X)-\delta_{Y,Z}\mu^*(X)$  & \ref{eqn:delid}  \\[5pt] \hline
\end{tabular}
\label{tab:sumdiff}
\end{table}

\begin{table}
\centering
\caption{This table answers the question: `Given $X$, how does one pick $Y$ and $Z$ to induce an opposite signed difference, and thus create the paradox?' Table~\ref{tab:definitions} contains a complete listing of all symbols used. The color coding used in the expression matches that used in Fig.~\ref{fig:oppositec}. The right column links back to the derivation of the expression.}
\begin{tabular}{lll}
Condition  & Expression  & Introducing Equation   \\[5pt] \hline
 &&\\[-1em]
$\triangle_Y \mu^{+}(X) < 0$ & ${\color{teal}{\mu^+(Y)}} < {\color{blue}{\mu^+(X)}}$ & \ref{eqn:fwdam}  \\[5pt] \hline
 &&\\[-1em]
$\nabla_Y \mu^{+}(X) < 0$ & ${\color{blue}{\mu^+(X)}} < {\color{teal}{\mu^+(Y)}}$ & \ref{eqn:condbackam} \\[5pt] \hline
 &&\\[-1em]
$\delta_{Y,Z} \mu^{+}(X) < 0$ & ${\color{teal}{\mu^+(Y)}} < {\color{orange}{\mu^+(Z)}}$ & \ref{eqn:amrep}  \\[5pt] \hline
 &&\\[-1em]
$\triangle_Y \mu^{*}(X) > 0$ & ${\color{teal}{\mu^*(Y)}} > {\color{blue}{\mu^*(X)}}$ & \ref{eqn:condfwdgm}  \\[5pt] \hline
 &&\\[-1em]
$\nabla_Y \mu^{*}(X) > 0$ & ${\color{teal}{\mu^*(Y)}} < {\color{blue}{\mu^*(X)}}$ & \ref{eqn:condbackgm} \\[5pt] \hline
 &&\\[-1em]
$\delta_{Y,Z} \mu^{*}(X) > 0$ & $\frac{{\color{teal}{\mu^*(Y)}}}{{\color{orange}{\mu^*(Z)}}}>1$ & \ref{eqn:gmrep} 
\end{tabular}
\label{tab:opposite}
\end{table}


\begin{table}[htbp]
\centering
\caption{Example of two cell lines, each with a different proportion of k-mers. The total number of proteins in each cell is the same (1059), but the transition dynamics are subtly different such that Cell {\color{orange}B}, compared to Cell {\color{blue}A}, has a larger proportion of {\color{teal}trimers}, a complex of three proteins, at the cost of a decreasing frequency of {\color{red}1-mers} and {\color{red}27-mers}. The last two rows show the equivalent probability (frequency) distribution. Note that while the frequency of the largest structures is very small, it will dominate visual perception in the actual images due to its disproportionate size. The transition model (Markov chain) is illustrated in Figure~\ref{fig:markov}-I. In terms of the chain, the transition $\frac{P[x_t=1, x_{t+1}=3]}{P[x_t=3, x_{t+1}=1]}$ in B is increased compared to A. 
Considering these as probability distributions of k-mers of two cell lines and repeatedly sampling from each, we have the resulting boxplots of the average cell-wise protein volumes (mean $k^{3}$ for each cell) in Fig.~\ref{fig:markov}-II. that demonstrate the paradoxical comparison (Eqn.~\ref{eqn:paradox}), i.e. $\mathbb{E}(A)>\mathbb{E}(B)$, yet $\mathbb{E}(\log.(A))>\mathbb{E}(\log.(B))$}

\begin{tabular}{lllllll}
\hline
\textbf{\# proteins / structure} & 1   & 3   & 9  & 27 & \begin{tabular}{@{}c@{}}Total \# \\ Structures\end{tabular} & \begin{tabular}{@{}c@{}}Total \# \\ Proteins\end{tabular} \\ \hline
\# structures in cell {\color{blue}A}  & 300 & 100 & 30 & 7  & \textbf{437} & \textbf{1059}  \\ \hline
\# structures in cell {\color{orange}B}                           & {\color{red}{240}} & {\color{teal}147} & 30 & {\color{red}4} & \textbf{437} & \textbf{1059}  \\ \hline
Frequency of structures in cell {\color{blue}A}  & 0.283 & 0.094 & 0.028 & 0.006  & \textbf{1} & \textbf{1}  \\ \hline
Frequency of structures in cell {\color{orange}B}                           & {\color{red}{0.227}} & {\color{teal}0.139} & 0.283 & {\color{red}0.003}  & \textbf{1} & \textbf{1} \\ \hline
\end{tabular}

\label{tab:markov}
\end{table}
\end{appendix}
%
%

\clearpage
\begin{acks}[Acknowledgments]
Ivan Robert Nabi is associated with the School of Biomedical Engineering, University of British Columbia, Canada. Sieun Lee is associated with the Mental Health and Clinical Neuroscience, School of Medicine, University of Nottingham, United Kingdom.
\end{acks}

\end{document}